\documentclass[fleqn,10pt]{wlscirep}

\usepackage{subcaption}
\usepackage{xcolor}

\usepackage{tabularx}

\usepackage{algorithm}
\usepackage{algpseudocode}

\title{Disentangling sources of influence in online social networks}

\author[1]{Matija Pi\v{s}korec}
\author[1]{Tomislav \v{S}muc}
\author[2,3,*]{Mile \v{S}ikic}
\affil[1]{Rudjer Boskovic Institute, Zagreb, Croatia}
\affil[2]{Faculty of Electrical Engineering and Computing, Zagreb, Croatia}
\affil[3]{Genome Institute of Singapore, A\*STAR, Singapore, Republic of Singapore}

\affil[*]{mile.sikic@fer.hr}


\begin{abstract} 
Information propagation in online social networks is facilitated by two types of influence - \emph{endogenous} (peer) influence that acts between users of the social network and \emph{exogenous} (external) that corresponds to various external mediators such as online news media. However, inference of these influences from data remains a challenge, especially when data on the activation of users is scarce. In this paper we propose a methodology that yields estimates of both endogenous and exogenous influence using only a social network structure and a single activation cascade. Our method exploits the statistical differences between the two types of influence - endogenous is dependent on the social network structure and current state of each user while exogenous is independent of these. We evaluate our methodology on simulated activation cascades as well as on cascades obtained from several large Facebook political survey applications. We show that our methodology is able to provide estimates of endogenous and exogenous influence in online social networks, characterize activation of each individual user as being endogenously or exogenously driven, and identify most influential groups of users.
\end{abstract}
\begin{document}

\flushbottom
\maketitle
\thispagestyle{empty}



Popularity of online social networks allows us to investigate dynamics of social interactions on a scale that was previously unattainable~\cite{Borge-Holthoefer2011, Kramer2014, Lewis2008, Karsai2014, Guille2013, DeDomenico2013, Najar2012, Yang2010}, while at the same time raising ethical concerns not previously encountered~\cite{Salganik2018,Kosinski2015}. One particular type of social interaction is an \emph{information cascade} - a spread of information between the users of a social network~\cite{Granovetter1978,Watts2002}. Information cascades are instrumental in investigating \emph{social influence}, which can be defined as the degree to which the behavior of individuals changes the behavior of their peers \cite{Aral2017}. Although mathematical modeling of social influence and information cascades is an active field of research in sociology for decades~\cite{Granovetter1978,Watts2002}, it only recently became technologically feasible to apply it to wide range of domains such as viral marketing \cite{Leskovec2007}, information diffusion \cite{Kimura2011}, behavior adoption \cite{Centola2010} and epidemic spreading \cite{Pastor-Satorras2001}.

The most commonly used information diffusion models were inspired by epidemiology which model how a disease spreads in a population~\cite{DKmodel,MTmodel,Hill2010}. However, their utility is sometimes hindered by their use of latent states which are unobservable in data. For this it is more appropriate to use Independent Cascade (IC) model~\cite{ICM} and Linear Threshold (LT) model~\cite{Granovetter1978,Kempe2003} which feature two observable states - \emph{active} and \emph{inactive} that denote whether an user was already exposed to the piece of information or not. These are popular for their simplicity that facilitates theoretical analysis~\cite{Narasimhan2015}, statistical inference from data~\cite{Goyal2010}, and can also be used as building blocks for more complex applications such as influence maximization~\cite{Wang2012}. However, there are several crucial differences between epidemic spreading and information diffusion~\cite{Centola2018}. Epidemic spreading is better modeled with simple contagion model where \emph{endogenous} factors play a dominant role, and the activation probabilities are independent of the neighborhood structure and the state of activated users in it. On the other hand, information diffusion is better modeled with complex contagion due to the common presence of \emph{exogenous} factors~\cite{Onnela2010} and more complex forms of endogenous influence which include various social reinforcement mechanisms such as reciprocity~\cite{Mahmoodi2018}, social feedback~\cite{Eckles2016} and homophily~\cite{Shalizi2011}. These additional factors are often neglected in modeling.

Presence of exogenous factors is particularly problematic as it confounds with the endogenous factors, and can be hard to differentiate using observational data alone~\cite{Aral2009}. Ideally, one would want to perform a study where exogenous influence is negligible \cite{Onnela2010}, but this is often not possible and exogenous influence has to be explicitly accounted for \cite{ArgolloDeMenezes2004, Myers2012, Karsai2016}. In fact, exogenous influence is instrumental for understanding the information spreading as information can propagate through multiple channels simultaneously, many of which are exogenous to the online social network itself - news media websites, direct communication via email and instant messengers, and even offline word-of-mouth transmission. In addition, external events such as political unrest~\cite{Borge-Holthoefer2011,Gonzalez-Bailon2011} and natural disasters~\cite{Lu2014} are often strong mediators of information cascades. These exogenous influences are usually not directly observable in the online social network itself, although they can be inferred from the available data. Understanding how endogenous and exogenous forces influence the information diffusion in online social networks could help us estimate to what extent are these vulnerable to manipulation by various interest groups such as organized individuals, news media and government agencies \cite{Lazer2018}.


In this paper we present a new methodology for estimation of endogenous and exogenous influence in online social networks. Our current model is conceptually similar to the unified model of social influence~\cite{Srivastava2015} which was shown to be generalization of many popular influence models, including complex contagion model~\cite{Karsai2014}, independent cascade model~\cite{Kempe2003} and generalized threshold model~\cite{Kempe2003}. In our previous work~\cite{Piskorec2017} we proposed a simpler method for inference of endogenous and exogenous influence that exploits statistical differences between the way the two types of influence act on users. The underlying assumption is that the endogenous influence is dependent on the current state of the social network and which users are already active or not, while the exogenous influence is independent on these. By incorporating these assumptions in a statistical model we can infer magnitude of endogenous and exogenous influence from empirical data.

Here, we develop a likelihood-based approach which is expressive enough to accommodate many different microscopic models of influence, and propose a maximum likelihood inference method to estimate the parameters. The inference problem is the following - given a single activation cascade and a friendship network between users, and assuming a particular form of endogenous influence, infer parameters of endogenous and exogenous influence and estimate magnitudes of these influences in time and on a global and user level. Similar attempts exist in literature, including peer and authority model~\cite{Anagnostopoulos2011} which, however, requires explicit modeling of \emph{authorities} responsible for exogenous influence, while in our case this is not necessary. Many of the other approaches rely on the availability of multiple activation cascades, while we use only one. Also, we use the social network structure, based on final state of activation cascade, directly in our inference rather than using it implicitly~\cite{Yang2010} or relying on a network statistic such as degree distribution~\cite{Brach2014}.

\begin{figure}
\begin{subfigure}[c]{0.48\textwidth}
    \begin{subfigure}[c]{\textwidth}
      \includegraphics[width=\textwidth]{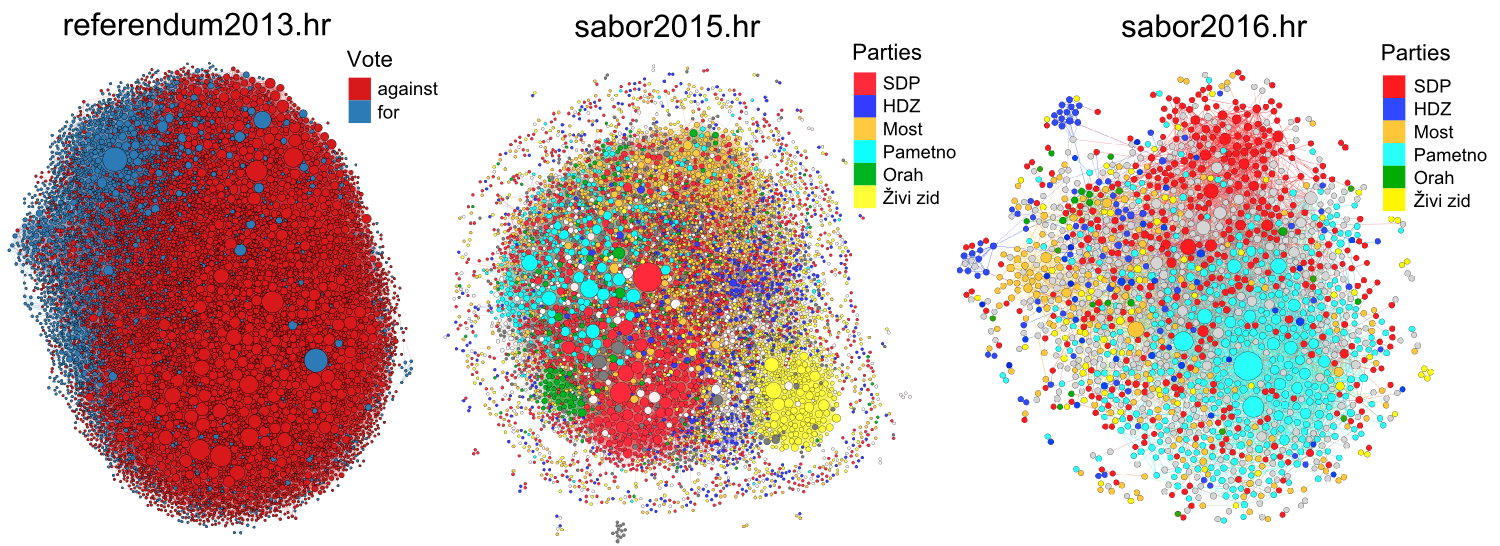}
      \caption{}
      \label{fig:networks_small}
    \end{subfigure}
    \vspace{0.5em}\\
    \begin{subtable}{\textwidth}
    \centering
    \setlength{\tabcolsep}{0.3em} 
    {\renewcommand{\arraystretch}{1.0}
    \begin{tabular}{llrl}
    \textbf{\scriptsize{Dataset}} & \textbf{\scriptsize{Time period}} & \textbf{\scriptsize{Users}} & \textbf{\scriptsize{Collected data}} \\
    \hline
    \texttt{\scriptsize{referendum2013.hr}} & \scriptsize{25.11. - 1.12.2013.} & \scriptsize{10175} & \scriptsize{friendships, demographics} \\
    \texttt{\scriptsize{sabor2015.hr}} & \scriptsize{2.11. - 8.11.2015.} & \scriptsize{6909} & \scriptsize{friendships, referral links} \\
    \texttt{\scriptsize{sabor2016.hr}} & \scriptsize{5.9. - 11.9.2016.} & \scriptsize{3818} & \scriptsize{friendships, referral links} \\
    \end{tabular}
    }
    \caption{\label{tab:dataset-info}}
    \label{tab:dataset}
    \end{subtable}
\end{subfigure}
\begin{subfigure}[c]{0.51\textwidth}
  \includegraphics[width=\textwidth]{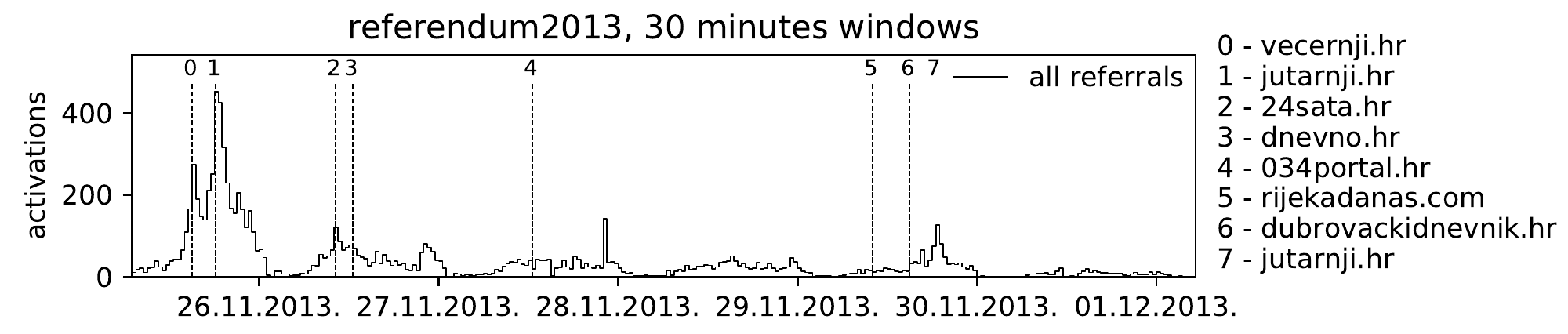}
  \includegraphics[width=\textwidth]{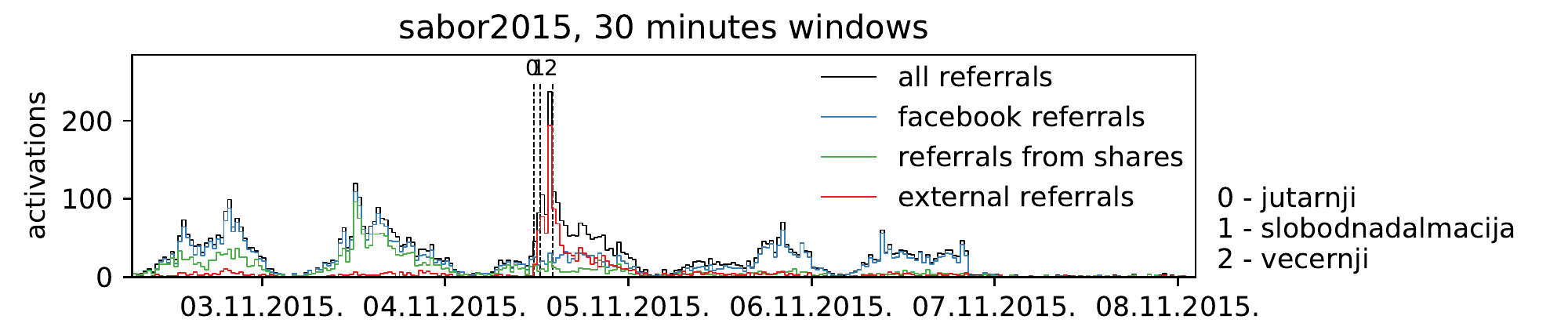}
  \includegraphics[width=\textwidth]{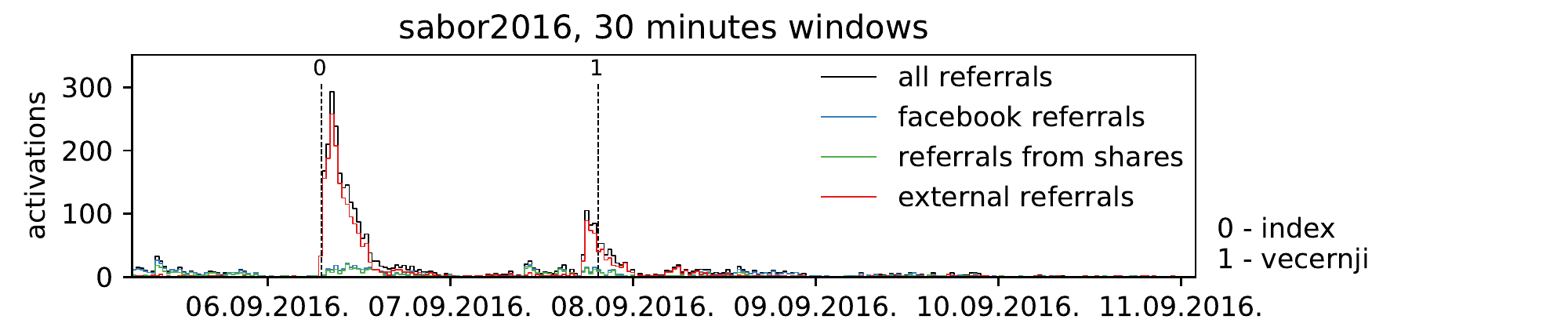}
  \caption{}
  \label{fig:activations}
\end{subfigure}
\caption{
\textbf{Descriptive statistics of collected datasets.} Visualization of Facebook friendship networks (Figure~\ref{fig:networks_small}) and registration time series (Figure~\ref{fig:activations}) of users who registered on three of our Facebook online survey applications: referendum2013.hr (11538 registered users), sabor2015.hr (6909 registered users) and sabor2016.hr (3818 registered users). Network nodes are colored according to the user's votes, and node sizes correspond to the number of their Facebook friends that also registered on the survey application. Clustering of users into communities based on votes shows a homophily effect - users are more likely to associate with other users that share their political preferences. This suggests a potential for endogenous influence. Time series are annotated with times of major news events which reported on our online survey application, and which are used as a proxy for exogenous influence. Collected data (table in Figure~\ref{tab:dataset-info}) include demographic information, friendships between users, and referral links through which users visited our applications. Time period refers to the period when surveys were active. Depending on whether these referral links originated within Facebook or some external website they could be used as indicators of endogenous and exogenous influence respectively. Time series for sabor2015.hr and sabor2016.hr datasets in Figure~\ref{fig:activations} are additionally separated based on the type of the referral links.
}
\label{fig:dataset}
\end{figure}

We evaluate our methodology on activation cascades collected via three online survey applications related to three distinct political events in Croatia (Figure~\ref{fig:dataset}). First survey, which is related to the referendum on the definition of marriage in 2013, we already used in our previous work~\cite{Piskorec2017}. Other two surveys are related to Croatian parliamentary elections in 2015 and 2016 and we collected them exclusively for this research. In all of our surveys the activation cascades are a series of user registrations through time. Surveys were active one week prior to actual elections and through them users were able to express their vote on the upcoming elections, see summary statistics for all users as well as for their online peers, and share the link to the survey through Facebook. Besides votes, we also collected Facebook friendship connections between all users that participated in our survey. In 2013 survey we also collected demographic data and in other two we obtained referral links through which users visited our survey website. These referral links originate either from Facebook, which indicates endogenous influence, or from some external website, which indicates exogenous influence. This classification of referral links served as a proxy for ground truth influence and allowed us to evaluate our inference method. During data collection we followed Facebook's privacy guidelines.

The main contributions of this paper are the following:  
(i) We collected data on social engagement of over 20 thousand Facebook users that participated on three distinct online political surveys. Datasets where users have to provide an informed consent to collect their data are usually much smaller, and so researchers have to rely on simulated datasets in order to validate their models.
(ii) We estimate magnitude of endogenous and exogenous influence in social networks by using only a single activation cascade of users and their friendship network. Most previous research relies on the availability of multiple information cascades and rarely tackles exogenous influence directly by either leaving it as an option\cite{Srivastava2015}, devising experiments where it is negligible \cite{Onnela2010} or simply treat it as a nuisance~\cite{Goyal2010}.
(iii) We show how can our methodology be used to estimate collective influence of various groups of users and characterize to what extent was their activation endogenously or exogenously driven. These estimates agree with both the simulated activation cascades and three realistic use cases where user's referral links served as a proxy for the ground truth labels on whether users were endogenously of exogenously activated.




\section*{Results}

Crucial components of our methodology are explicit microscopic models of endogenous and exogenous influence with which we expand the Independent Cascade (IC) model. We then use these models in a log-likelihood function which gives us probability of observing particular activation cascade as a function of the model's parameters. Formulating our inference problem in a probabilistic way allows us to optimize for the maximum likelihood parameters and to estimate the magnitude of endogenous and exogenous influence. We apply our methodology on several simulated and empirical activation cascades in order to characterize the activation of users as being more endogenously or exogenously driven. The simulated case is easier because we know both the functional form and the parameters of the model that generated simulated information cascade, which allows us to perform evaluation in a straightforward manner. For the empirical cases we use three Facebook datasets obtained from an online political survey applications. In the end we estimate collective influence of three groups of users - those who registered by following link from within Facebook, those that registered by following link from an external website, and those that followed a link from a Facebook advertisement.

\subsection*{Models of endogenous and exogenous influence}

We assume that an activation of an user in an online social network is mediated by two influences (Figure~\ref{fig:influence}): (i) endogenous influence $p_{peer}$ which depends on the network structure and users that are already active or not, and (ii) exogenous influence $p_{ext}$ which is modeled as a time dependent random variable and is constant across all users. An additional assumption is that parameters of endogenous influence are constant throughout the period of observation, while parameters of exogenous influence may change in time. Both sets of parameters are equal for all users. This allows us to use a very simple model for the exogenous influence - a single probability of activation $p^{(i)}_{ext}(t)$ which is equal for all inactive users $i$ at each specific time step, although it can change in time. Instead of parameterizing $p^{(i)}_{ext}(t)$ with a suitable closed form, we chose to evaluate it at each time step independently~\cite{Myers2012}.

For the endogenous influence we choose two commonly used Independent Cascade (IC) models: (i) Susceptible-infected (SI) model $p^{(i)}_{SI}(t)$ and (ii) Exponential decay (EXP) model $p^{(i)}_{EXP}(t)$. IC models are an example of \emph{simple contagion} - activation of users happens due to a direct influence of one of their peers, independently of the rest of the system, including the neighborhood structure and which other users are active or not. EXP model has an added condition that peers that activated recently carry more influence than the ones that activated farther away in time, which is commonly incorporated in endogenous influence models \cite{Takaguchi2013, Karimi2013}. 

Probability of endogenous activation for user $i$ at time interval $[ t - \Delta t, t]$ under the SI model is defined as follows:

\begin{equation}
p^{(i)}_{SI}(t) = 1 - \prod_{j \in N^{(i)} \text{active at } t} (1-p_{0}) = 1 - (1-p_{0})^{a_{i}(t)}
\label{eq:SI}
\end{equation}
where $N^{(i)}$ is a set of peers of user $i$, $a_{i}(t)$ designates how many of them are active at time $t$, and $p_{0}$ is a probability of user $i$'s being activated by each of its peers. Assumption of the SI model is that probability of activating one's peers does not change in time, so once user is activated, every subsequent step he has the same probability $p_{0}$ of activating any of his peers. This assumption is more appropriate in epidemiological setting, from where SI model originated, than in information propagation setting where we would expect the influence to decay in time. This could be achieved by adding a parameter for influence decay, which leads us to the EXP model:

\begin{equation}
p^{(i)}_{EXP}(t) = 1 - \prod_{j \in N^{(i)} \text{active at } t} (1-p_{0} e^{-\lambda (t-t_{j})})
\label{eq:EXP}
\end{equation}
where $t_{j}$ is the time of activation of user $j$. $p_{0}$ and $\lambda$ are parameters of endogenous influence which define the shape of exponential decay of influence, with $p_{0}$ being the probability of user $j$ activating user $i$ at time $t = t_{j}$ and $\lambda$ being the half-decay of influence. Both SI and EXP models feature independent cascades - each individual user can independently activate any of his peers. However, in social contagion it is more realistic to add a requirement of multiple interactions for the activation. This effectively models social reinforcement mechanism which is a known driving force for product adoption \cite{Onnela2010}. One of the simplest examples of such \emph{complex contagion} models is the \emph{threshold model} where the probability of endogenous activation is related to the number of already active peers $N^{(i)}$ of user $i$. We define one such threshold model in the Equation S11 of the Supplementary and show that it can also be effectively incorporated into our inference methodology.


We now define a likelihood function $\mathcal{L}$ which gives us probability of observing data $D$ (network and activation times) at a particular time $t$ given some functional forms for endogenous and exogenous influence $p_{peer}$ and $p_{ext}$. Due to typically small probabilities involved in these processes we actually use log-likelihood for maximum likelihood estimation of parameters, where product of probabilities is replaced with the sum of log-probabilities:

\begin{equation}
\begin{split}
\log \mathcal{L}(D;p_{peer},p_{ext},t) = 
&\sum_{i \in \text{activated at } [t - \Delta t, t]} \log (1-(1-p^{(i)}_{peer}(t))(1-p_{ext}(t))) + \\
&c(t) \sum_{i \in \text{inactive at } t} \log ((1-p^{(i)}_{peer}(t))(1-p_{ext}(t)))
\end{split}
\label{eq:log-likelihood}
\end{equation}

First term on the right-hand side quantifies the agreement for the users that \emph{did} activate in a given time period $[t-\Delta,t]$, as this had to be due to either endogenous or exogenous influence. Second term quantifies the agreement for the users that \emph{did not} activate up to time $t$, neither through endogenous nor through exogenous influence. The time enters our inference \emph{only} through the activation time of users and is used in two ways - i) to determine which users were active or inactive in time window $[t - \Delta t, t]$ (Equation~\ref{eq:log-likelihood}), and ii) to calculate endogenous influence decay in EXP model (Equation~\ref{eq:EXP}). However, in principle it is possible to use a \emph{temporal network} where friendship connections between users change in time. This would have to be encoded into the expression for endogenous influence $p_{peer}$. We can remove explicit dependence on time $t$ by evaluating $\mathcal{L}$ nonparametricaly - at each time increment $\Delta t$.

One issue still needs to be addressed - on which users does the exogenous influence actually acts? We know that our friendship network does not contain \emph{all} possible users, and so the true number of yet inactive users is probably much larger than what we actually observe. This \emph{observer bias} could lead to the overestimation of the exogenous influence as we approach the end of the activation cascade and the number of eventually observed inactive users decreases towards zero, while the true number of inactive users which could possibly activate (but do not during our observation period) stays large. We correct for this by artificially increasing the part of our log-likelihood which is responsible for inactive users by factor $c(t) = 1 + \alpha (N_{all}/N_{inactive}(t))$, where $N_{all}$ is the number of all users in the social network, and $N_{inactive}(t)$ the number of all yet users inactive users at time $t$ (more details in Section S7 of Supplementary).

\subsection*{Maximum likelihood inference for endogenous and exogenous influence}

\begin{figure}
\begin{subfigure}[c]{0.24\textwidth}
  \includegraphics[width=\textwidth]{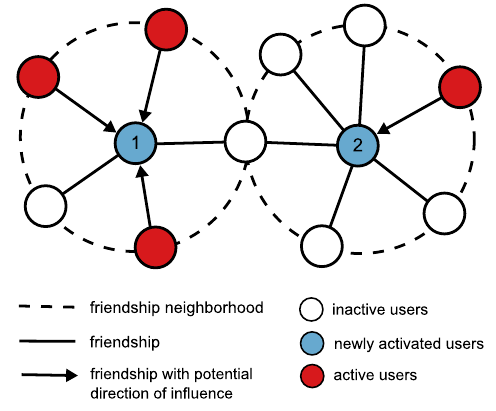}
  \caption{}
  \label{fig:network-peer-external-influence}
\end{subfigure}
\hspace{2em}
\begin{subfigure}[c]{0.70\textwidth}
  \includegraphics[width=\textwidth]{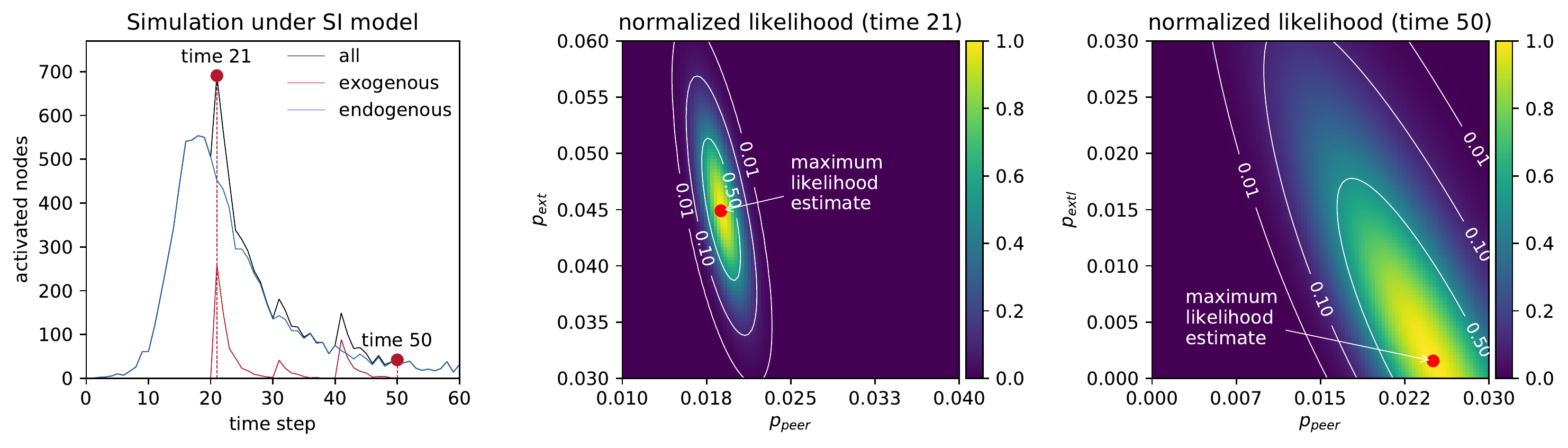}
  \caption{}
  \label{fig:visualization_of_likelihood}
\end{subfigure}
\caption{
\textbf{Maximum likelihood inference of endogenous and exogenous influence.} Our assumption is that information propagation in an online social network is mediated by two types of influence - endogenous (peer) which acts between the users of the social network and exogenous influence which is external to it (Figure~\ref{fig:network-peer-external-influence}). The estimated endogenous influence on the newly activated user $i=1$ should be higher because more of his peers are already active, as compared to user $i=2$. Figure~\ref{fig:visualization_of_likelihood} shows the normalized likelihood function (similar to Equation~\ref{eq:log-likelihood} which shows log-likelihood) at two distinct time steps in the simulated activation cascade using SI model for endogenous influence. SI model features only two parameters at each time step - parameter of endogenous influence $p_{\text{peer}}$ ($p_0$ in Equation~\ref{eq:SI}) and a parameter of exogenous influence $p_{\text{ext}}$. Shape of the likelihood function suggests that these two parameters are correlated as each provides part of the explanation for the observed data, and if one is weaker the other most compensate. Also, when we have more data (time 21) the shape of the log-likelihood function is more concentrated than when we have less (time 50), resulting in more confident estimates. In this simulation we are estimating parameters of endogenous and exogenous influence at each time step separately, which corresponds to the initialization stage of our actual inference procedure which we use in simulated (Figure~\ref{fig:simulation}) and empirical (Figure~\ref{fig:real_datasets_results}) case. In our full inference procedure we infer a single set of endogenous influence parameters for the whole observation period instead of having a separate estimate for each time step like in this example (more details in Methods section and in the Section S4 of the Supplementary). Here we are using a truncated Newton algorithm~\cite{Nash2000} for optimizing a log-likelihood function in order to obtain a maximum likelihood solution, although in practice any suitable optimization method could be used.
}
\label{fig:influence}
\end{figure}

We want to compute a single set of endogenous influence parameters for the whole period and a separate set of exogenous influence parameters for every time window. Our assumption is that endogenous influence parameters do not change over time, but that exogenous do. A direct way to do this is to perform a joint optimization of a log-likelihood that contains a single set of endogenous influence parameters and a separate set of exogenous influence parameters for each time window $[t + \Delta t]$. Our log-likelihood would then be $t+1$-dimensional in the case of SI model, and $t+2$-dimensional for the EXP model - $t$ parameters of exogenous influence for each time window we are considering in our inference plus the parameters of endogenous influence ($p_0$ for SI model and $(p_0,\lambda)$ for EXP model). This makes the number of parameters proportional to the number of time windows, which makes a joint optimization of log-likelihood unfeasible. Instead, we use an alternating method~\cite{Myers2012} where we alternatively fix either endogenous influence parameters or exogenous influence parameters and optimize the other until both values converge. In addition, we never optimize all of the $t$ parameters of the exogenous influence jointly but do it one by one. This yields a nonparametric estimate for exogenous influence, meaning that we have a separate estimate of exogenous influence $p_{ext}(t)$ at each time step $t$. Although the number of parameters we have to infer is still proportional to the number of time windows we are considering in our inference, this strategy is much more efficient then joint inference and provides reliable estimates even though there is no formal guarantee that the estimates will actually converge. However, in our experiments we did not experience any problems with the convergence. Figure~\ref{fig:visualization_of_likelihood} shows the initialization step of the alternating procedure on a simple simulated activation cascade, where parameters for endogenous and exogenous influence are inferred separately for each time step $t$.

Using efficient optimization routines allows our method to scale to networks of over 10000 users with resolution of 100 time steps. In our experiments we use a truncated Newton algorithm~\cite{Nash2000} for maximum likelihood estimation, although in principle any suitable optimization algorithm could be used (more details in Methods section and in Section S4 of the Supplementary). Total number of users activated due to endogenous and exogenous influence (in Figures \ref{fig:simulation_EXP} and \ref{fig:real_datasets_results}) is calculated through the \emph{exogenous responsibility} measure (Equation~\ref{eq:responsibility}) which is derived from the inferred parameters and quantifies the extent to which is each user's activation is due to endogenous or exogenous influence. This estimate is normalized with the total number of user activations in a given time interval, which is an observable quantity. 

\subsection*{Inference of endogenous and exogenous influence on simulated data}

\begin{figure}
\begin{subfigure}[c]{0.66\textwidth} 
  \includegraphics[width=\textwidth]{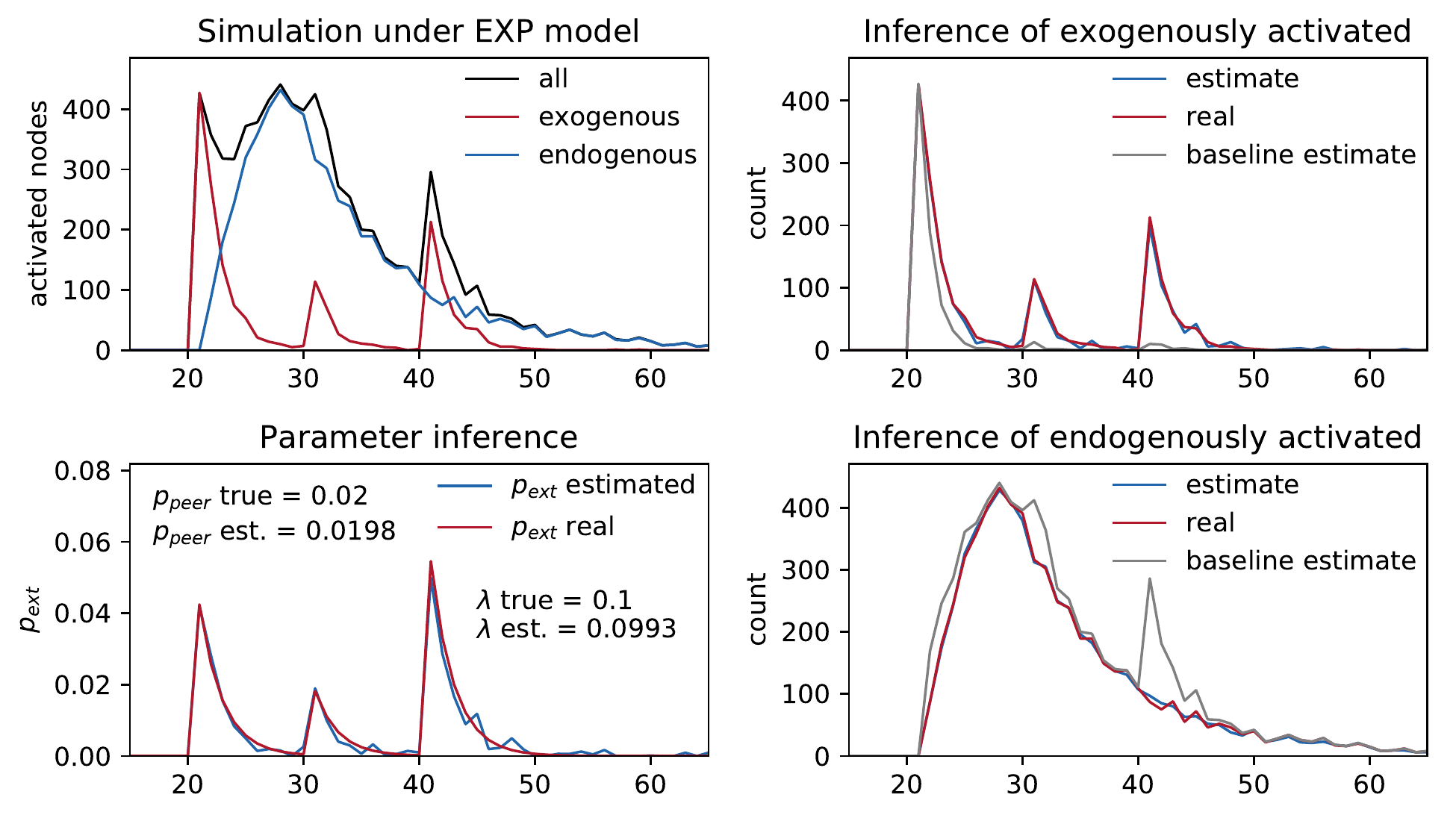}
  \caption{}
  \label{fig:simulation_EXP}
\end{subfigure}
\begin{subfigure}[c]{0.35\textwidth} 
  \includegraphics[width=\textwidth]{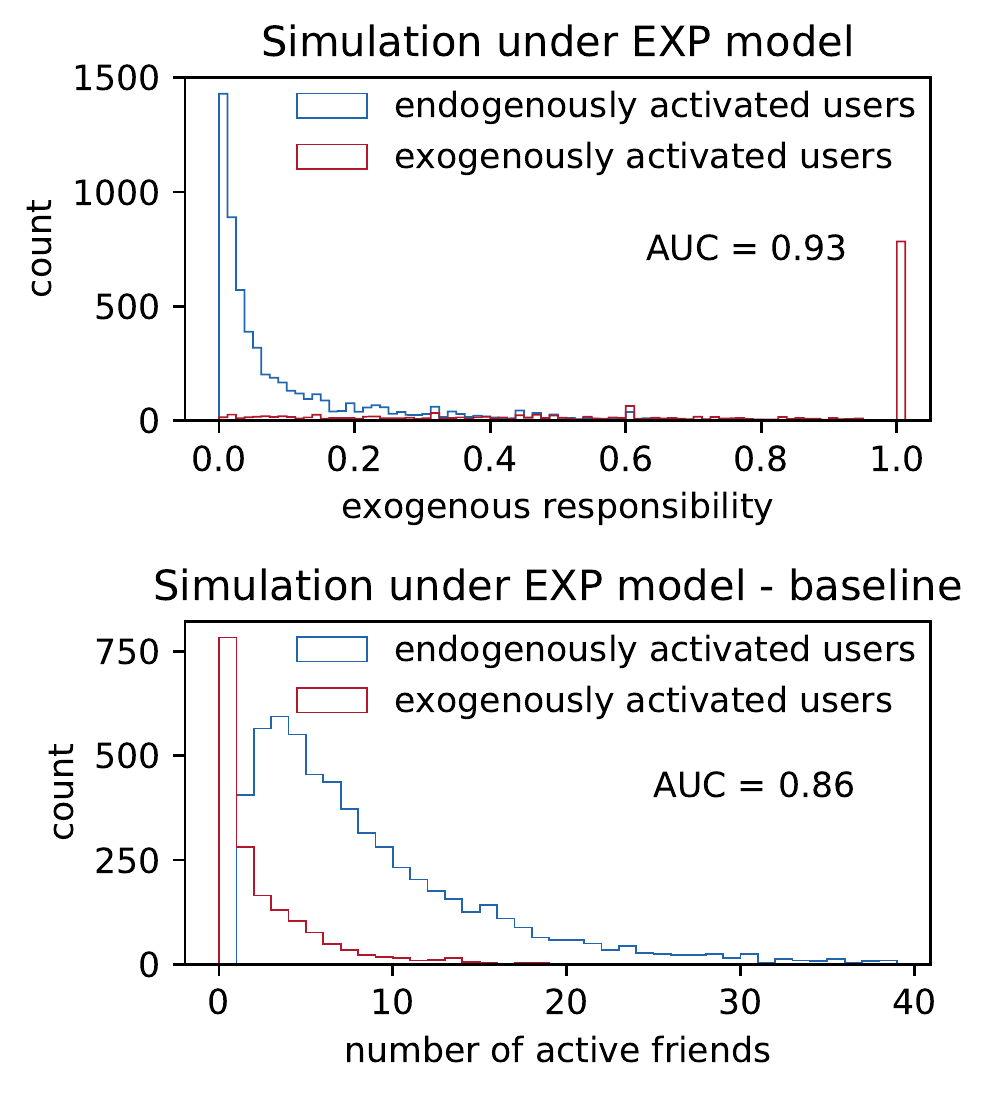}
  \caption{}
  \label{fig:simulation_activation_types_EXP}
\end{subfigure}
\caption{
\textbf{Inference on a simulated activation cascade.} We use our methodology to infer which users activated due to endogenous or exogenous influence in a simulated activation cascade following exponential decay (EXP) endogenous influence model. In real world applications only total number of activated users (black line) is actually observed, along with the friendship network between users (Figure~\ref{fig:simulation_EXP}). We use a configuration model of referendum2013 social network to make our results reproducible even without the whole empirical network. We see that our measure is able to differentiate absolute numbers of endogenously and exogenously activated users throughout the whole cascade period and to correctly infer the parameters of endogenous influence - $p_{peer}$ and $\lambda$, and exogenous influence $p_{ext}(t)$ for every time period $t$. We also infer activation type for each user individually by using the \emph{exogenous responsibility} measure $R^{(i)}(t)$ (Equation~\ref{eq:responsibility}) as shown on Figure~\ref{fig:simulation_activation_types_EXP} and achieve AUC of $0.93$. We compare this with the baseline method where, instead of exogenous responsibility, we use number of active peers at the time of activation. A special case of this baseline is where we consider users without any active peers as exogenously activated, which is a baseline that we use in Figure~\ref{fig:simulation_EXP}. This baseline method underestimates the exogenously activated users towards the end of the observation period, which is due to the fact that more and more users are active and it is increasingly likely that at least one of the peers is active by chance alone. On Figure~\ref{fig:simulation_activation_types_EXP} we show a histogram of the number of active peers and compare it with exogenous responsibility to demonstrate that no reasonable threshold could not serve as a classification measure, which is also confirmed with a relatively low AUC score of $0.86$. The results for SI endogenous influence model are similar and are available in Figure S6 in the Supplementary.
}
\label{fig:simulation}
\end{figure}

Our simulations are designed to approximate, as well as possible, the conditions in which real data were collected. However, instead of using one of the empirical social networks which we collected, we decided to simulate on a configuration model of referendum2013 Facebook friendship network so that our results are reproducible using only a degree sequence, which is a much more compact and anonymous representation in comparison to the whole empirical network. Configuration model of a network preserves the number of connections each user has, but these connections are permuted randomly across all users. This destroys mesoscale structures such as communities, but is still preferable to other permutation methods where either times of activation are permuted (destroying order of activity) or connections themselves are permuted between the users (destroying degree distribution by changing it to binomial) \cite{Holme2015}. The simulation starts with a small number of active users and progresses in discrete steps following one of the endogenous influence models (Equations~\ref{eq:SI}-\ref{eq:EXP}). Figure~\ref{fig:simulation} shows the results using the EXP model (Equation~\ref{eq:EXP}) for endogenous influence. At three distinct times we also simulate an exponentially decaying exogenous influence which acts equally on all inactive users. This resembles a typical situation when a distinct exogenous information source activates some of the users \cite{Crane2008}, which we also observe in our dataset (Figure~\ref{fig:activations}). However, our methodology works equally well for other shapes of exogenous influence (Figures S8 and S9 in the Supplementary). Using just the activation times of all users and their friendship network we are able to estimate the parameters of the assumed endogenous and exogenous influence models as well as the absolute number of users activated predominantly due to the one or the other. In addition, using a measure of \emph{external responsibility} (Equation~\ref{eq:responsibility}) we are able to infer, for user, the extent to which endogenous or exogenous influence was responsible for activation. Instead of using a single threshold to classify users we calculated the whole receiver operating characteristic (ROC) curve and the corresponding area under the curve (AUC) score to evaluate the performance (Figure~\ref{fig:simulation_activation_types_EXP}). We compare our method to a simple baseline commonly used in previous work~\cite{Myers2012,Gomez-Rodriguez2011} where an activation is considered exogenous if activated user had no other active peers at the time of the activation. However, as more and more users becomes active, it becomes increasingly likely that a user is connected with at least one other active user by pure chance. This underestimates the number of users activated by exogenous influence and consequently underestimates overall exogenous influence. We obtain similar results (Figure S6 in the Supplementary) for the SI endogenous influence model and an additional threshold model we define in the Equation S11 of the Supplementary. The inference itself is fast and scales well to networks of over ten thousand users (Section S6 in the Supplementary).

\subsection*{Inference of endogenous and exogenous influence on empirical datasets}

In order to investigate social interactions between users of a large online social network we developed three online surveys that use Facebook API for collection of data. Surveys were related to three distinct political events in Croatia: 1) \texttt{referendum2013.hr} for referendum on definition of marriage, 2) \texttt{sabor2015.hr} for parliamentary elections in 2015, and 3) \texttt{sabor2016.hr} for parliamentary elections in 2016. Figure~\ref{fig:dataset} shows the collected friendship networks between Facebook users and the number of registrations in 30-minute intervals for each of the survey applications during a week preceding the actual elections. Table in Figure~\ref{tab:dataset-info} shows summary statistics for each of the datasets. The referral links provide information whether each user followed a link originating from a post on Facebook which indicates endogenous influence, or some external website reporting on our survey which indicates exogenous influence. We use this information to evaluate our estimates of endogenous and exogenous influence acting on users. More details on the datasets and the methodology of data collection is available in the Methods section and Sections S1 and S2 of the Supplementary.

Figure~\ref{fig:real_datasets_results} shows the results of applying our inference methodology to estimate the magnitude of endogenous and exogenous influence during these three activation cascades. In this experiment we use the EXP model as endogenous influence model because it performed best on average over all three empirical datasets, with and without correction for the observer bias. The results for other models are included in Figures S12 and S13 of the Supplementary. As our methodology operates in discrete time (Equation~\ref{eq:log-likelihood}) we discretized the activation times of users into 30 minutes time intervals to determine which users were active or inactive during each specific interval. Considering the duration of the data collection for each of the surveys, this corresponds to $333$ time intervals for referendum2013 dataset, $327$ intervals for sabor2015 dataset and $328$ intervals for sabor2016 dataset. Each user that registered on one of the online survey application using his Facebook credentials is considered \emph{activated} in the given time period. The referral link from which we visited the website of the survey application will be used as a proxy of endogenous and exogenous influence - referral links from Facebook are considered as endogenous and those from external websites as exogenous. We later use this information for evaluation of our methodology.

We estimate magnitudes of endogenous and exogenous influence and characterize each user as being endogenously or exogenously activated. We use the AUC score to evaluate the predictive performance of our inferred model on sabor2015 and sabor2016 datasets for which we had data on referral links from which users visited our survey application. This served as a proxy for ground truth labels which we needed for calculating the AUC scores. The purpose of the model is to estimate the magnitude of endogenous and exogenous influence on each given user, given available data and provided that underlying assumptions of our statistical methodology are satisfied. Similar as in simulated experiments, we compare our methodology with a baseline method that simply estimates the number of exogenously activated users as all those who did not have any active peers at the time of their own activation, and again we observe that it underestimates the number of exogenously activated users, especially near the end of the observation period. Our estimates of endogenously activated users (Figure~\ref{fig:real_datasets_results}) closely resemble the true number of users activated by following another user's share, which is the strongest indication of endogenous influence we have. On the other hand, it might seem that our method overestimates exogenously activated users by declaring many of the users originating from Facebook as exogenously activated. However, relying on Facebook referrals alone is not a reliable proxy for endogenous activation, as many users might be activated through other means of indirect communication available through Facebook - by following an advertisement, or by directly visiting a Facebook page of the survey application.

We observe that the magnitude of exogenous influence increases as we approach the end of the activation cascade period. This effect is due to the fact that we only observe the friendship network of users that eventually registered on our application, which is only a small subset of the whole Facebook network. However, one of our assumption is that exogenous influence acts uniformly on all users in the friendship network, not just the subset of them, and this manifests in the increased exogenous influence as the activation cascade approaches the size of the network. This \emph{observer bias} can be corrected by adding a correction factor $c$ to our log-likelihood function (Equation~\ref{eq:log-likelihood}), which is regulated with parameter $\alpha$. The results of applying the correction term on the empirical data are shown on Figure~\ref{fig:real_datasets_results}, while more detailed experiments are available in Figure S5 of the Supplementary). However, because less and less users got activated near the end of the observation period this observer bias does not influence our final estimates by much. However, we still believe that correction is warranted and useful, especially for estimates near the end of the observation period, and in other use cases where observation period is shorter and observer bias might be more pronounced.

\begin{figure}
  \centering
  \includegraphics[width=\textwidth]{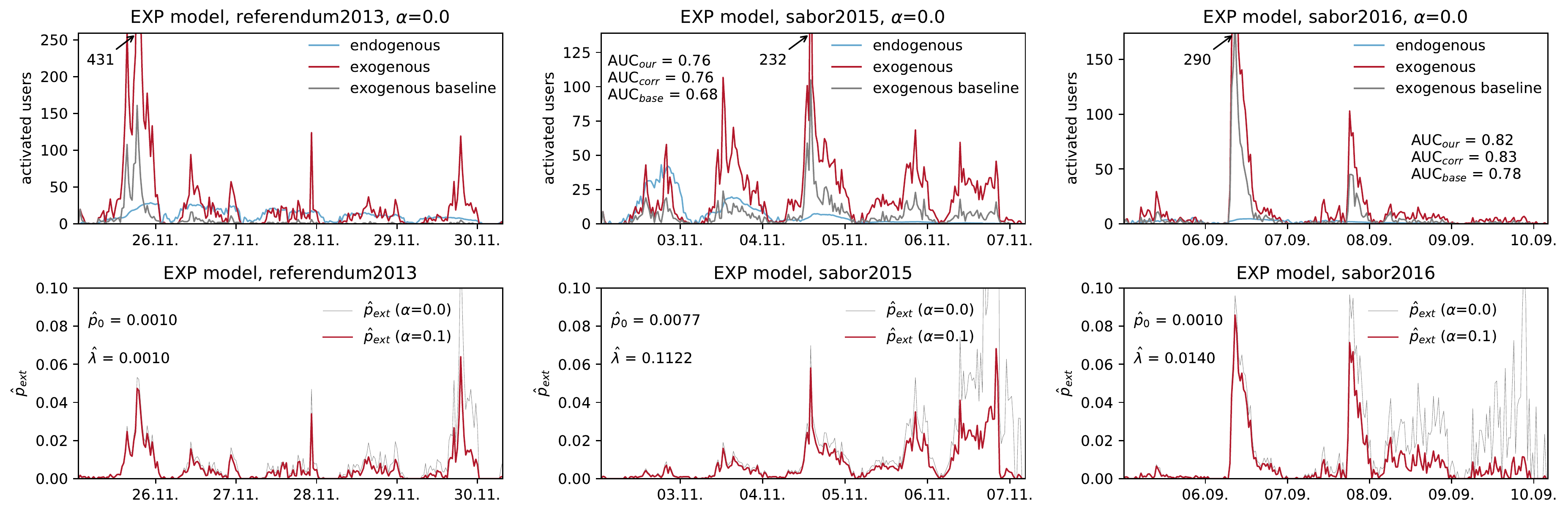}
\caption{\textbf{Inference on Facebook activation cascades with EXP model.} Inference of endogenous and exogenous influence on activation cascades derived from referendum2013, sabor2015 and sabor2016 online survey applications, with EXP model as assumed endogenous influence model. The results for the SI endogenous influence model are in Figures S12 and S13 of the Supplementary. On the bottom panels we see the effect of correction for the observer bias ($\alpha=0.1$) as compared to no correction ($\alpha=0$) - it reduces the overestimate of exogenous influence near the end of the observation period. AUC scores for using exogenous responsibility as a measure for classifying users into endogenously and exogenously activated ($AUC_{our}$) for datasets where we have information on referral links for evaluation - sabor2015 and sabor2016, are $0.76$ and $0.82$ respectively. This is higher then those achieved with a baseline measure of number of active friends, which are $0.68$ and $0.78$ for sabor2015 and sabor2016 datasets respectively. A more direct comparison with the baseline is available in Figure S14 of the Supplementary. Facebook referrals alone are not discriminating enough as there are multiple possible ways by which Facebook users might reach our application, including visiting the webpage of our application directly or through an advertisement, both which are more similar to exogenous rather than endogenous influence.}
\label{fig:real_datasets_results}
\end{figure}

For evaluation (Figure~\ref{fig:real_datasets_results}) we again calculate the corresponding AUC score which uses exogenous responsibility measure $R^{(i)}(t)$ (Equation~\ref{eq:responsibility}) to classify users into endogenously and exogenously activated. The achieved AUC scores for our method ($AUC_{our}$) for sabor2015 and sabor2016 datasets are $0.76$ and $0.82$ respectively. This is higher than the baseline measure which uses number of active peers at the time of activation which achieves AUC scores ($AUC_{base}$) of $0.68$ and $0.78$ for the sabor2015 and sabor2016 datasets respectively. Using exponential decay model for endogenous influence allows us to calculate the half-decay of endogenous influence which is $10.1$ hours for the sabor2015 dataset. This value is consistent with what we could expect, as it means that endogenous influence diminishes to a fraction of a value in the span of a day or two and requires influx of new users to keep it sustained.

\subsection*{Collective influence}

\begin{figure}
\begin{subfigure}[c]{0.29\textwidth}
  \includegraphics[width=\textwidth]{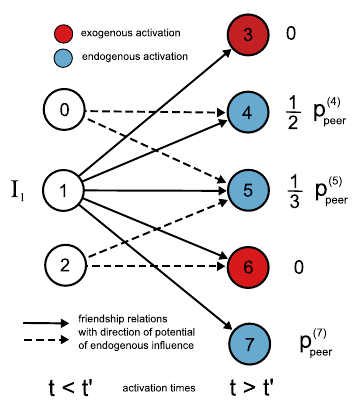}
  \caption{}
  \label{fig:individual_influence}
\end{subfigure}
\hspace{1em}
\begin{subfigure}[c]{0.68\textwidth}
  \includegraphics[width=\textwidth]{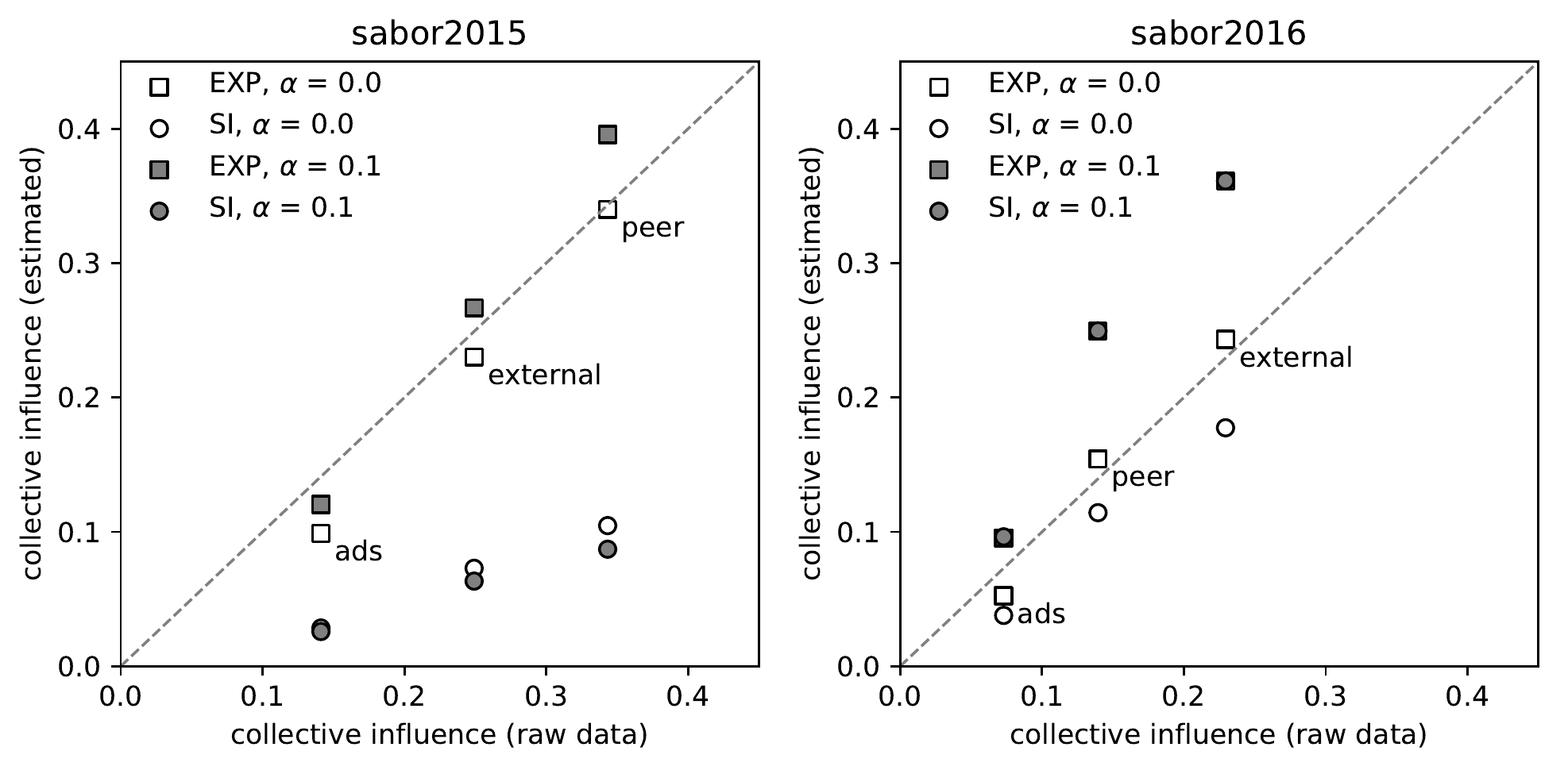}
  \caption{}
  \label{fig:mean_individual_influence_by_group_comparison}
\end{subfigure}
  \caption{\textbf{Individual and collective influence.} In the example on Figure~\ref{fig:individual_influence} we estimate influence $I_{1}$ of user $i=1$ as the extent to which he is responsible for endogenous activation $p^{(j)}_{peer}$ of all of his peers $j=\{3,4,5,6,7\}$ which activated after him. Only three of his peers $j=\{4,5,7\}$ activated due to endogenous influence, but he has to share part of this claim with two users $i=\{0,2\}$ which are their shared peers. The total individual influence for user $i=1$ in the above example is $I_{1} = 1/2 p^{(4)}_{peer} + 1/3 p^{(5)}_{peer} + p^{(7)}_{peer}$. Type of activation (endogenous or exogenous) for each user can be estimated with our methodology or taken from raw data by using referral links from which users visited our application, in which case $p^{(j)}_{peer}$ simply takes values $0$ or $1$. Figure~\ref{fig:mean_individual_influence_by_group_comparison} shows comparison of influence estimates obtained from our methodology and raw data for different groups of users - those activated due to endogenous (peer) influence, exogenous (external) influence and advertisements (ads). Ads are similar to exogenous influence as they are targeting large number of users independent of their friendship connections, but within the Facebook social network itself.}
  \label{fig:raw_estimate_comparison}
\end{figure}

Once we characterized activation of each user as being endogenously or exogenously driven, we can estimate the extent to which each user contributed to the activation of its peers by excluding the portion of the influence attributed to exogenous factors. We do not have a deterministic propagation path for our activation cascade - we do not know who influenced whom directly, so we cannot deterministically incorporate influence of all users in a transitive manner\cite{Teng2016}. Nevertheless, our measure of influence simply incorporates all \emph{possible} endogenous propagation paths to estimate an influence for each user (Figure~\ref{fig:individual_influence} and Equation~\ref{eq:individual_influence}). If we then average this influence over a group of users we get their \emph{collective influence}. Instead of using our estimates of endogenous and exogenous activation for each user we could also estimate influence directly from data by using the referral links from which users visited our application. Figure \ref{fig:raw_estimate_comparison} shows the comparison of our methodology with estimates of influence obtained from raw data for different groups of users that activated due to: endogenous factors, exogenous factors, advertisements. Our question was: Which channel of communication is the most influential, that is, recruits users with higher collective influence? The results of our experiments (Figure~\ref{fig:mean_individual_influence_by_group_comparison}) on two datasets for which we had data on referral links, shows no clear pattern of influence. Different groups of users are more influential depending on the dataset. However, regardless of the model of endogenous influence (SI or EXP) our estimates are robust and are proportional to the ones obtained from raw data. It is important to emphasize again that our methodology does not use any information on referral links or external influence whatsoever, but rather infers this from the dynamics of the user activations. More details is available in Section S5 of the Supplementary.

\section*{Discussion}

Unlike traditional survey methods where data is manually entered either by a respondent or experimenter~\cite{Dhand2018}, online social networks provide an opportunity to collect much larger amounts of data on user activity. However, due to their nature they provide challenges to experimental design~\cite{Walker2015}. Observational studies without explicit consent are regularly performed within companies for marketing purposes, which is regulated by company's privacy policy, and in some cases this research can be used for academic purposes~\cite{Eckles2016}. Still, academic publication of such research could raise ethical concerns~\cite{Kramer2014,Verma2014}. On the other hand, conducting a study where explicit consent is mandatory heavily restricts the amount of data that can be collected, even when researchers have a direct access to the whole online social network and are in position to present their experiment automatically to the large number of users. For example, a study from Aral and Walker~\cite{Aral2012} on a sample of 1.3 million Facebook users managed to collect responses of only 7730 users. However, major publicized events such as elections and referendums can serve as catalyzers for mobilizing users. Users are usually willing to participate in a study if through it they receive an information or a service which they perceive as valuable and which could not be easily obtained in some other way. 

Despite inherent difficulties in collecting data, we decided to conduct several online surveys using our own web applications and Facebook's API, which allowed us to collect activation cascades and friendship connections of over 20 thousand users in total. Although computational social science  is in its infancy, with standards and practices still taking shape, we tried to keep the privacy of the users and follow current recommended ethical practices~\cite{Kosinski2015, Salganik2018}. Conducting a survey through an online social network means that the recruitment happens organically from person to person as a form of snowball sampling and not through some unbiased randomized procedure, so it's the most eager persons that are recruited first. Number of mobilized users mostly depends on highly connected and willing individuals, that mobilize less wiling users. This effect might easily dominate the one from mass media~\cite{Rutherford2013}. 


Using this data we demonstrate how to estimate exogenous and endogenous influence using only information on the friendship connections between users and a single activation cascade which corresponds to the times of user registration. Our methodology exploits the different ways of how exogenous and endogenous influence propagate - endogenous influence propagates between users and as such is dependent on the friendship structure, while exogenous influence acts uniformly on all users regardless of the social network structure. Our method is not able to reconstruct an exact propagation pathway, as these inevitably include pathways external to the particular online social network as well as pathways that are inherently unobservable such as word-of-mouth communication. Still, our method is able to give a probabilistic estimates of these two influences given minimal assumptions. Any additional information on the activation cascade or the social network could be included in our methodology, most probably along the lines of the unified model of social influence~\cite{Srivastava2015}. The advantage of such likelihood-based approaches is that inference is performed in a probabilistically-consistent manner, instead of relying on aggregated statistics to choose among competing models of influence~\cite{Leskovec2008}. The availability of efficient numerical solvers means our method can easily scale to large networks of over 10000 users. Computational scalability was already addressed for the unified model~\cite{Popa2017}, however, only for the modeling and not for inference. Our methodology could be applied for  characterizing the types of influence in information spreading, for example the role of external factors in the fake news spreading occurring over online social networks such as Facebook or Twitter~\cite{Bovet2019}. Also, there might also be applications outside the domain of social networks as the paradigm of endogenous and exogenous effects could be applied in the wider context of dynamical systems modeling~\cite{ArgolloDeMenezes2004}.

Our methodology suffers from several limitations, which also indicate potential paths for future research. First, we do not elucidate the mechanisms by which endogenous and exogenous influence arise. The form of the endogenous influence is predefined, and choosing between several possible candidates is possible. In our case, we evaluate different endogenous influence models by their prediction on empirical data, but other methods are possible, including information-theoretic approaches. Second, we assume exogenous influence acts equally on all users, and that parameters of endogenous influence are equal for all users. This was necessary in our case because we only have one activation cascade available for inference~\cite{Karsai2016}, and without imposing additional constraints our statistical inference would be infeasible~\cite{Yoshikawa2010,Du2014}. In cases where multiple activation cascades are available, it should be possible to relax these assumptions and allow for different values of endogenous and exogenous influence parameters for various groups of users. Third, we do not try to correct for the confounding effect arising from unobserved or observed characteristics of users. For example, it is expected that users respond differently to influences, both exogenous and endogenous, from entities that share their political orientation as compared to those that do not. Again, including additional parameters in our model would increase the uncertainty of our estimates.

\section*{Methods}


\subsection*{Alternating method for inference}

Our two main assumptions during statistical inference are: (i) both endogenous and exogenous influence are equal for all users at any given time, and (ii) endogenous influence does not vary in time while exogenous influence does. This leads us to the inference algorithm where we seek a single set of parameters for the endogenous influence $p_{peer}$ and a set of parameters for the exogenous influence $\{p_{ext}\}_{t}$ for each time step $t$. This would make the dimensionality of our log-likelihood proportional to the number of time steps we use for inference, which would be hard to optimize numerically. Instead, we use an \emph{alternating} method~\cite{Myers2012} where we alternatively fix either $p_{peer}$ or $\{p_{ext}\}_{t}$ and optimize for the other. The inference procedure is the following:

\begin{enumerate}
\item Estimate maximum likelihood values for $p_{peer}$ and $p_{ext}$ for every time window separately.
\item Fix $p_{ext}$ to values obtained from each time window and estimate a single maximum likelihood value $p_{peer}$ for the whole period.
\item Fix $p_{peer}$ to the single value obtained for the whole period and estimate maximum likelihood value $p_{ext}$ for every time window separately.
\item Repeat from step 2 until estimates for $p_{peer}$ and $p_{ext}$ converge.
\end{enumerate}

An actual maximum likelihood estimation in steps 1 to 3 is performed with a truncated Newton algorithm that is Hessian-free and uses conjugate gradients to iteratively compute parameter updates~\cite{Nash2000}, although in principle any suitable optimization algorithm could be used (more details in Section S6 of the Supplementary). A full pseudocode for this alternating method is available in Section S4 of the Supplementary.

\subsection*{Inference of activation types}

Because our model gives us probabilities for endogenous and exogenous activation for each user individually, we can use this information to estimate activation type for each of the users. For this we define a single measure of \emph{exogenous responsibility} $R^{(i)}$ which quantifies to what degree is an activation of user $i$ due to the exogenous (external) influence: 

\begin{equation}
R^{(i)}(t) = \frac{p_{\text{ext}}(t)}{p_{\text{ext}}(t) + p^{(i)}_{\text{peer}}(t)}
\label{eq:responsibility}
\end{equation}

Where $t$ is the time of activation of user $i$. Values close to zero indicate dominating endogenous influence, and values close to one indicate dominating exogenous influence. An extreme value of zero is achieved for users who activated during time when there was no exogenous influence acting in the network. An extreme value of one is achieved for users who, at the time of their activation, did not have any active peers. Note that it is not possible for both $p_{\text{ext}}(t)$ and $p^{(i)}_{\text{peer}}(t)$ to be $0$, and consequently that the value of responsibility is undefined, because that would mean the activation of this user is evaluated as \emph{impossible} by our model in Equation~\ref{eq:log-likelihood}. In principle, we could also use pure activation probabilities $p^{(i)}_{\text{peer}}$ or $p^{(i)}_{\text{ext}}$ as measures of influence, but experiments on simulated data showed that exogenous responsibility is the most sensible (more details in Supplementary Information).

\subsection*{Individual and collective influence of users}

Our assumption is that each user is, to some extent, responsible for endogenous activation of all of his peers that activated after him. This influence extends beyond user's immediate peers. However, as we do not have a deterministic activation path (we do not know who shared information with whom) it is not straightforward to transitively incorporate influence from far away users as it is usually done \cite{Teng2016}. This is why we express the influence $I^{(i)}$ of user $i$ (Equation~\ref{eq:individual_influence}) as the extent to which user $i$ is responsible for activation of his peers $j$: 

\begin{equation}
I^{(i)} = \sum_{j \in N^{(i)}} \frac{I^{(i \to j)}}{\sum_{m \in N^{(j)}} I^{(m \to j)}} p^{(j)}_{peer}(t_j)
\label{eq:individual_influence}
\end{equation}

Where $I^{(i \to j)}$ is the fraction of the endogenous influence that user $i$ can claim for user $j$. In our case we define it as $I^{(i \to j)} = 1$ if $i$ and $j$ are peers, and $0$ otherwise. This means that all user's are credited equally for the activation of their peers, regardless of how far away in time they themselves activated. For an alternative formulation which involves time see Equation S8 in the Supplementary. As shown on Figure \ref{fig:individual_influence}, each user can claim part of the peer activation probability $p^{(j)}_{peer}(t_j)$ for each of his peers $j$ that activated after him $t_{i} < t_{j}$. As we do not have a deterministic activation path, this is really just a potential for responsibility and so the user has to share part of his claim to $I^{(i \to j)}$ with all other $m$ peers of $j$. For the SI model we can set this to $1$, meaning that we consider all peers equally responsible regardless of the time of their activation. Each user would then be assigned $1/m$ of the peer activation probability $p^{(j)}_{peer}$ for each of his peers that activated after him, where $m$ is the number of user's $j$ peers that activated before him. For the EXP model we can weight this with the times of activation - users can claim larger part of the influence for peers that activated close in time to their own activation (more details in Section S5 of the Supplementary). The collective influence for a group of users $G$ is just an average influence of all users in the group $1/|G|\sum_{i \in G} I^{(i)}$.

\subsection*{Evaluation}

Instead of using a single threshold for the exogenous responsibility to classify users into endogenously and exogenously activated we calculate the entire receiver operating characteristic (ROC) curve and associated area under the curve (AUC) score. This allows us to compare different endogenous influence models regardless of the chosen threshold. In order to calculate the ROC curve and AUC score we also need some sort of a gold standard label for each user, for which we use referral links available for sabor2015 and sabor2016 datasets. Depending on the referral link we classify users in one of the three categories (Figure \ref{fig:real_datasets_results}): (i) strong endogenous influence for users whose referral link originates from a Facebook share, (ii) potential endogenous influence for users whose referral link originates from Facebook and (iii) strong exogenous influence for users whose referral link originates from an external web site. Users who do not have a referral link are considered as unknown. For the purpose of evaluation we consider users from category (i) as endogenously activated and users from category (iii) as exogenously activated.

\subsection*{Data collection}

Our online survey applications were actually web applications which used Facebook Graph API~\cite{GraphAPI} for authentication of users. Some sort of user authentication was necessary to prevent multiple voting. In addition, Graph API allowed us to collect Facebook friendship relationship between users registered on our application. In addition, with referendum2013.hr we collected basic demographics information such as age and gender and with other three applications we collected referral links through which users visited our web application. These we collected through our own web server which hosted the survey application, not the Graph API. Before users registered they had to accept the privacy policy of the application which was in complete alignment with with Facebook's platform policy~\cite{FacebookPolicy2017} (more details in Section S2 of the Supplementary). Facebook's Graphs API assigns application-specific ID's to each user, so it is not possible to associate users from different datasets. After they registered users were able to see summary voting statistics of their friends as well as for all registered users. These statistics were displayed \emph{after} the user cast his vote in order to minimize the influence on his choice. We also provided an additional incentive to share the link to the application through Facebook and other social media by displaying to each user a number of users which registered to the application after following the referral link from their share, and comparing this to other users.

\subsection*{Code availability}

Instructions and the code to reproduce the results from this paper are available upon a request after signing the data access agreement on \url{https://goo.gl/forms/IxINFkeBSJpDuzRv2}. Facebook online survey applications through which we collected referendum2013 and sabor2015 datasets are also available on public Github repositories: \url{https://github.com/devArena/referendum2013.hr}, and \url{https://bitbucket.org/marin/sabor2015.hr}. More information is available in Sections S1 and S2 of the Supplementary.

\bibliography{bibliography-main}

\section*{Acknowledgements}

The work is supported in part by the Centre of Excellence project ``DATACROSS'', co-financed by the Croatian Government and the European Union through the European Regional Development Fund - the Competitiveness and Cohesion Operational Programme (KK.01.1.1.01.0009). We would like to thank the people with whom we had fruitful discussions and who helped in various stages of manuscript preparation and experimental design: Nino Antulov-Fantulin, Vinko Zlati\'{c} and Sebastian Krausse. Also, we greatly appreciate the effort of people who actively collaborated in the development of the Facebook online survey applications with which we collected the data: Bruno Rahle, Iva Miholi\'{c}, Tomislav Lipi\'{c}, Vedran Ivanac, Matej Mihel\v{c}i\'{c} and Mladen Marinovi\'{c}.

\section*{Author contributions statement}

M.P. and M.\v{S} conceived the experiments. M.P. performed the experiments, developed the methodology and wrote the manuscript. All authors analyzed the results and reviewed the manuscript. 

\subsection*{Data availability}

Due to Facebook's privacy policy \url{https://developers.facebook.com/policy} we are not allowed to publicly release any Facebook-derived data, including personal information and friendship relations between our users. Friendship networks and registration times needed to reproduce the results of this paper are available upon a reasonable request and only after signing the data access agreement stating that (i) you will only use the dataset for the purpose of reproducing and validating the results of our study; (ii) you shall not attempt to deanonymize the dataset or in any other way compromise the identity or privacy of users contained in it; and (iii) you shall not further share, distribute, publish, or otherwise disseminate the dataset. The access agreement is available at \url{https://goo.gl/forms/IxINFkeBSJpDuzRv2}.

\section*{Additional information}

\textbf{Competing interests}: Part of the data was provided by Oraclum Intelligence Systems Ltd.


\end{document}


\flushbottom
\maketitle
\thispagestyle{empty}

\tableofcontents

\section{Descriptive analysis of the collected datasets}

As a part of our research we collected three large datasets on Facebook users that registered on one of our online political survey applications (tables \ref{tab:dataset} and \ref{tab:sabor2015}) related to three distinct political events in Croatia: 1) referendum on the definition of marriage (raferendum2013 dataset), 2) parliamentary elections in Croatia in 2015 and 2016 (sabor2015 and sabor2016 datasets). We already used referendum2013 dataset in one of our previous research~\cite{Piskorec2017}. Depending on the survey we collected different data - for referendum2013.hr application we collected demographics data but without referral links, while for all the subsequent applications we collected referral link but without demographics data. In our case referral links are much more useful because they allow us to evaluate our methodology for estimation of endogenous and exogenous influence. On the other hand, demographic data could be used to build more complex model of influence by correcting for the potential confounder variables. However, as we decided to restrict ourselves to a simple model of influence that only takes into consideration friendship connections between users and their times of registration, we decided not to collect demographic starting from the sabor2015 survey application.

\begin{table}[ht]
\centering
\begin{tabular}{llrll}
\textbf{Dataset} & \textbf{Time period} & \textbf{Users} & \textbf{Collected data} & \textbf{URL of application} \\
\hline
\texttt{referendum2013.hr} & 25.11. - 1.12.2013. & 10175 & friendships, demographics & \tiny{\url{https://github.com/devArena/referendum2013.hr}} \\
\texttt{sabor2015.hr} & 2.11. - 8.11.2015. & 6909 & friendships, referral links & \tiny{\url{https://github.com/matijapiskorec/sabor2015.hr}} \\
&&&&
\tiny{\url{https://bitbucket.org/marin/sabor2015.hr}} \\
\texttt{sabor2016.hr} & 5.9. - 11.9.2016. & 3818 & friendships, referral links & \\
\hline
\end{tabular}
\caption{\label{tab:dataset-info} \textbf{Expanded version of the summary statistics for the collected datasets.} Time period refers to the period when surveys were active, which is typically one week prior to the actual elections. Friendships and demographics were collected through the Facebook Graph API, while referral links were collected from our own web server that was hosting the survey application. Source codes of the first two survey applications are published on Github and are freely available.}
\label{tab:dataset}
\end{table}

\begin{figure}
\begin{subfigure}[c]{0.49\textwidth}
  \includegraphics[width=1.0\textwidth]{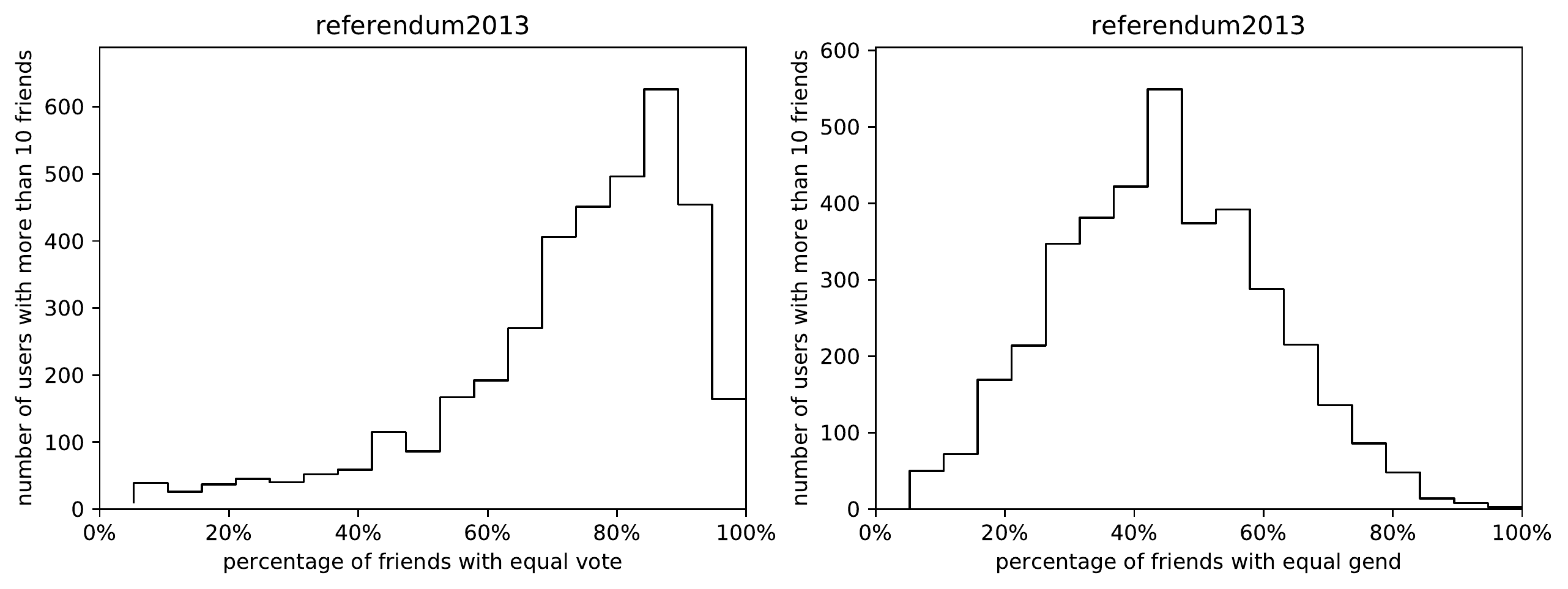}
  \caption{}
  \label{fig:homophily}
\end{subfigure}
\begin{subfigure}[c]{0.49\textwidth}
  \includegraphics[width=1.0\textwidth]{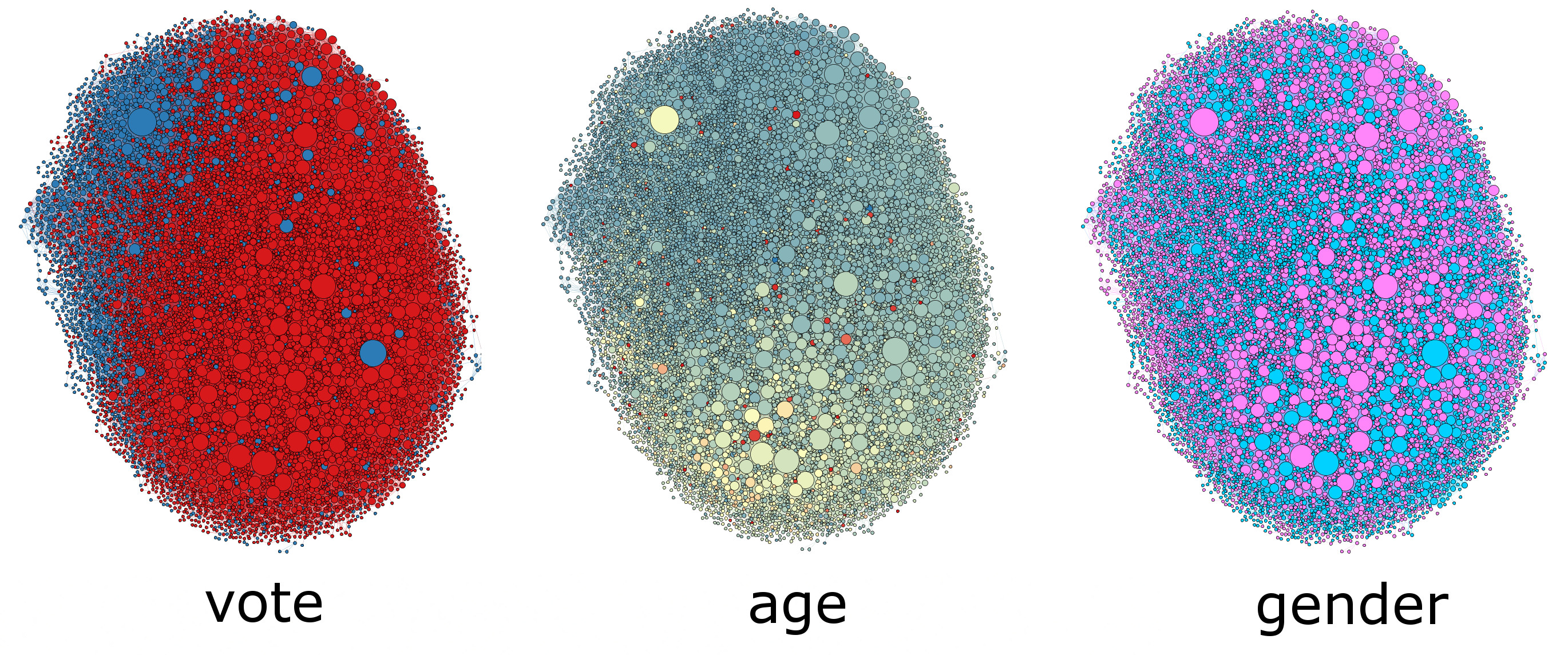}
  \caption{}
  \label{fig:networks}
\end{subfigure}
\caption{
\textbf{Homophily and the Facebook friendship networks for the referendum2013 dataset.} Figure~\ref{fig:networks} shows Facebook friendship network for referendum2013 dataset colored by three attributes - vote (blue for ``for'' votes and red for ``against'' voters), age (pale blue for for voters bellow 30 years of age, pale yellow for middle age voters and orange-red for voters above 50 years of age), and gender (pink for female voters and blue for male voters). Size of the nodes correspond to the their degree - the number of friends they have. Figure~\ref{fig:homophily} shows homophily with respect to gender and with respect to votes - users are much more likely to friend other users that share their voting preferences, while they are equally likely to friend users of both gender. Homophily with respect to age is visible on the network of users (also on Figure~\ref{fig:networks}). 
}
\label{fig:networks_homophily}
\end{figure}

\begin{figure}
\centerline{\includegraphics[width=\textwidth]{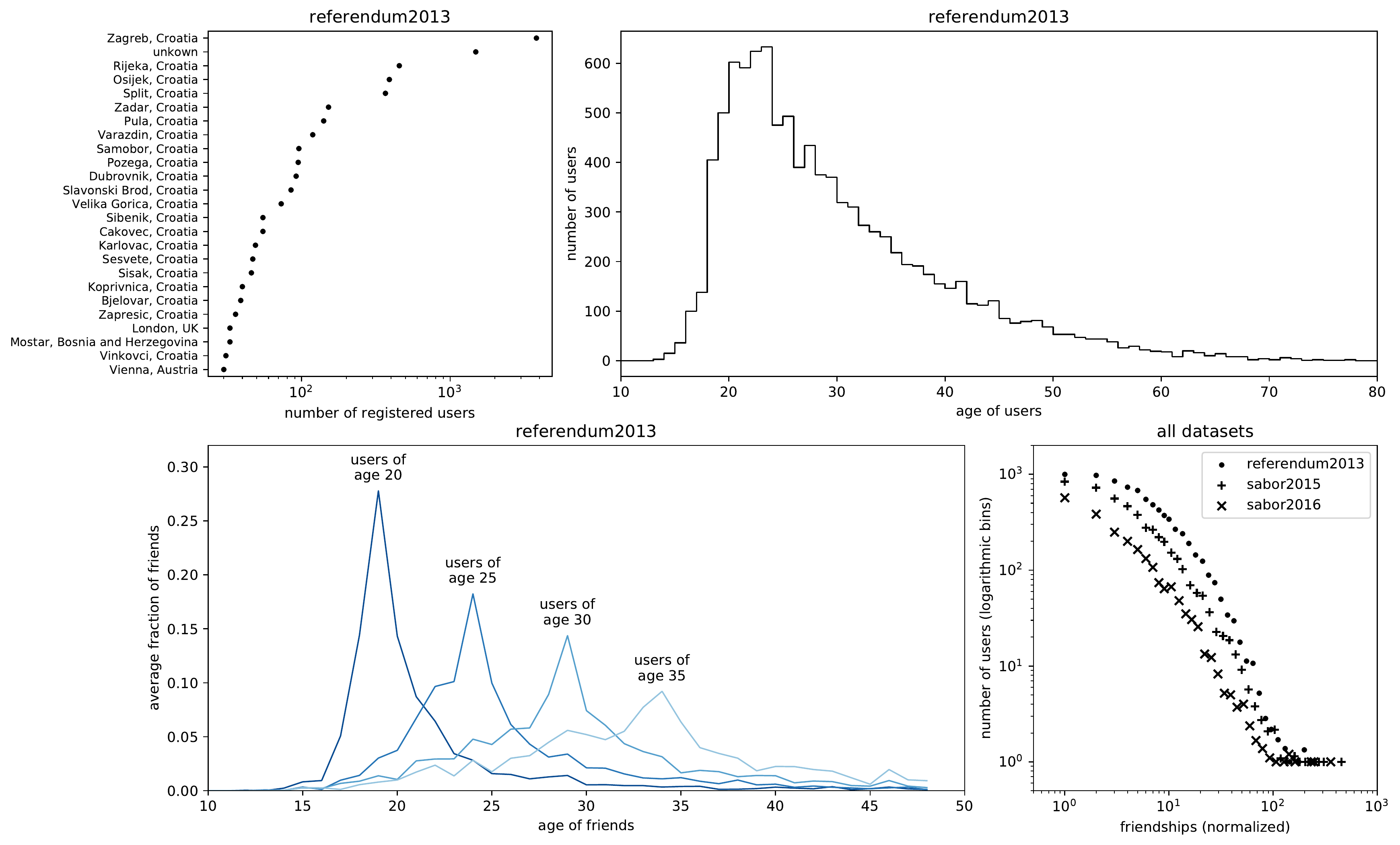}}
\caption{\textbf{Exploratory analysis of the referendum2013 dataset.} Self-reported locality information (top left) shows that majority of users come from Zagreb, Croatia. We did not restrict participation on the survey based on the location as we were also interested in the opinions of Croatia's citizens living abroad. Language of the survey was Croatian so we believe this served as the most effective filter. Age distribution (top right) and age of friends for users of different ages (bottom left) show that the average user is much younger than expected from the population census. However, we were more interested in obtaining a representative sample of Facebook population rather than obtaining a representative sample of the population itself, and these statistics are qualitatively consistent with the ones obtained from the whole Facebook network \cite{Ugander2011}. Degree distribution of the number of friends for all three datasets (bottom right) is also consistent from what is expected in online social networks.}
\label{fig:descriptive_statistics}
\end{figure}

Because we had collected demographic information of users only for the referendum2013 dataset, we decided to perform more detailed exploratory analysis on that dataset. Exploratory analysis of the friendship network of voters immediately reveals large homophily with respect to votes (Figure \ref{fig:homophily}) and age (Figure \ref{fig:descriptive_statistics}). Homophily with respect to votes is the strongest, with majority of users having $80\%$ or more friends who voted the same as they did, which indicates potential presence of endogenous influence between users. On the other hand, homophily with respect to gender is almost nonexistent, with users being equally likely to friend users of both gender. These statistics are consistent with study performed on a much larger Facebook friendship networks~\cite{Ugander2011}. Table \ref{tab:sabor2015} shows first few lines of the sabor2015 dataset which stores the information on the information cascade. More information on the collected data and our data sharing policy is available in the main paper.

\begin{table}[ht]
\centering
\begin{tabular}{lllllll}
\textbf{User id} & \textbf{Time login*} & \textbf{Time share} & \textbf{Referrer id**} & \textbf{Referrer class**} & \textbf{Friend count} & \textbf{Election list id} \\
\hline
0 & 4798 & -1 & -1 & facebook & 363 & 37 \\
1 & 5684 & 5691 & -1 & facebook & 88 & 8 \\
2 & 2099 & -1 & 3145 & facebook & 485 & 37 \\
3 & 4073 & -1 & 4816 & facebook & 861 & 4 \\
4 & 5471 & -1 & -1 & facebook & 108 & 8 \\
5 & 1106 & -1 & -1 & facebook & 53 & 4 \\
\hline
\end{tabular}\\
\raggedleft * used in inference, ** used in evaluation
\caption{\textbf{First few lines of the sabor2015 data on user sessions.} All users are identified by unique survey-specific id's which could not be traced back to their actual Facebook identities. Times are recorded as minutes from reference time which is usually a short time before survey applications went online. Time of login corresponds to the time of the first login by the user to our survey application. We sometimes refer to this as the user's \emph{registration time}. If a user shared a link to the survey application through his Facebook account, this time is also recorded as ``Time share'', otherwise it is $-1$. If a user visited the survey application by following a referral link from another user's share, this user's id is also recorded, otherwise it is $-1$. A fact that user followed a share is a strong indication of potential endogenous influence between users, and we supplement this with a general category of a referrer - Facebook if referral originated from Facebook (not necessarily from a share), and a specific name of an external website (usually an online news site) if referral originated from there. Information on the total number of Facebook friends (``Friend count'') and vote on the survey (``Election list id'' in the case of sabor2015 dataset) we do not use explicitly in the inference. Along with this information on user sessions, our inference method also uses a social network dataset on the Facebook friendship relations between users. These are available in GML format and as an edge list.}
\label{tab:sabor2015}
\end{table}

\section{Facebook application for collecting data and survey methodology}

We developed online survey applications as a separate web pages which used Facebook Graph API~\footnote{\url{https://developers.facebook.com/docs/graph-api}} to allow Facebook users to register with their Facebook accounts. The survey was hosted on an independent server with its own database, with the authorization as a crucial component which allowed us to uniquely track the identity of users. Otherwise it would be impossible to know whether a particular person registered multiple times. The Graph API allowed us to retrieve, for each user, all Facebook friends that also registered on our application. From this data we constructed a friendship graph used in our inference methodology. It also allowed us to retrieve additional demographic data on users: their age, gender, hometown. We collected these for our first application referendum2013.hr, but later decided to collect only the friendship network and the referral links for our subsequent applications. It should be noted that users still had to give and explicit permission for each of these variables in order for us to collect them. Permissions were given through the Facebook API's interface. As we do not use demographic data in our inference we decided not to collect it for all subsequent survey applications after referendum2013. The unnecessary collection of demographic data entails a potential security risk due to the possibility of deanonymization - identifying specific users in the dataset. Deanonymization strategies often rely on combining the user data obtained from different sources, for example different Facebook applications or even different social networks. Any piece of information which is shared between these sources, demographics being the most notable one, increases the risk of partial or even full deanonymization. From our side, only information that is shared both by Facebook and us is the subset of the friendship network. Registration times of the users and the referral links from which they visited our web page are specific for our application and as such cannot aid much in the deanonymization attempt. We should note that Facebook also changed its Graph API several times in the last couple of years due to security concerns. Most notably, it is not possible anymore to collect data on user's friends, only an absolute number of friends for each user and a friendship connection toward all other users that are also users of the same application. Also, each Facebook application is given an unique identifying number for each users which are not shared among applications. This prevents owners of multiple applications to directly cross-identify users between different applications.

\begin{figure}
  \centering
  \begin{subfigure}[c]{\textwidth}
    \includegraphics[width=1.0\textwidth]{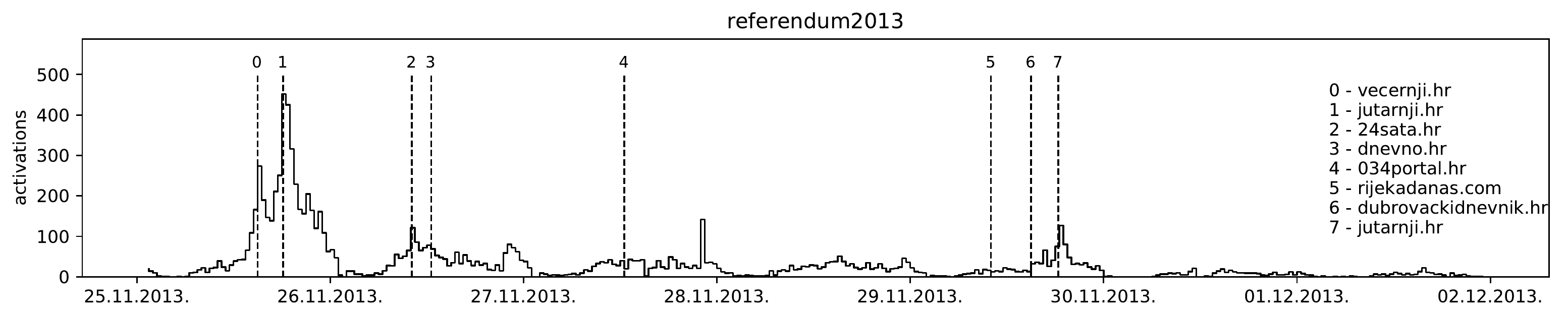}
    \caption{}
    \label{fig:activations_referendum2013}
  \end{subfigure} \\
  \begin{subfigure}[c]{\textwidth}
    \begin{subfigure}[c]{0.33\textwidth}
      \includegraphics[width=0.9\textwidth]{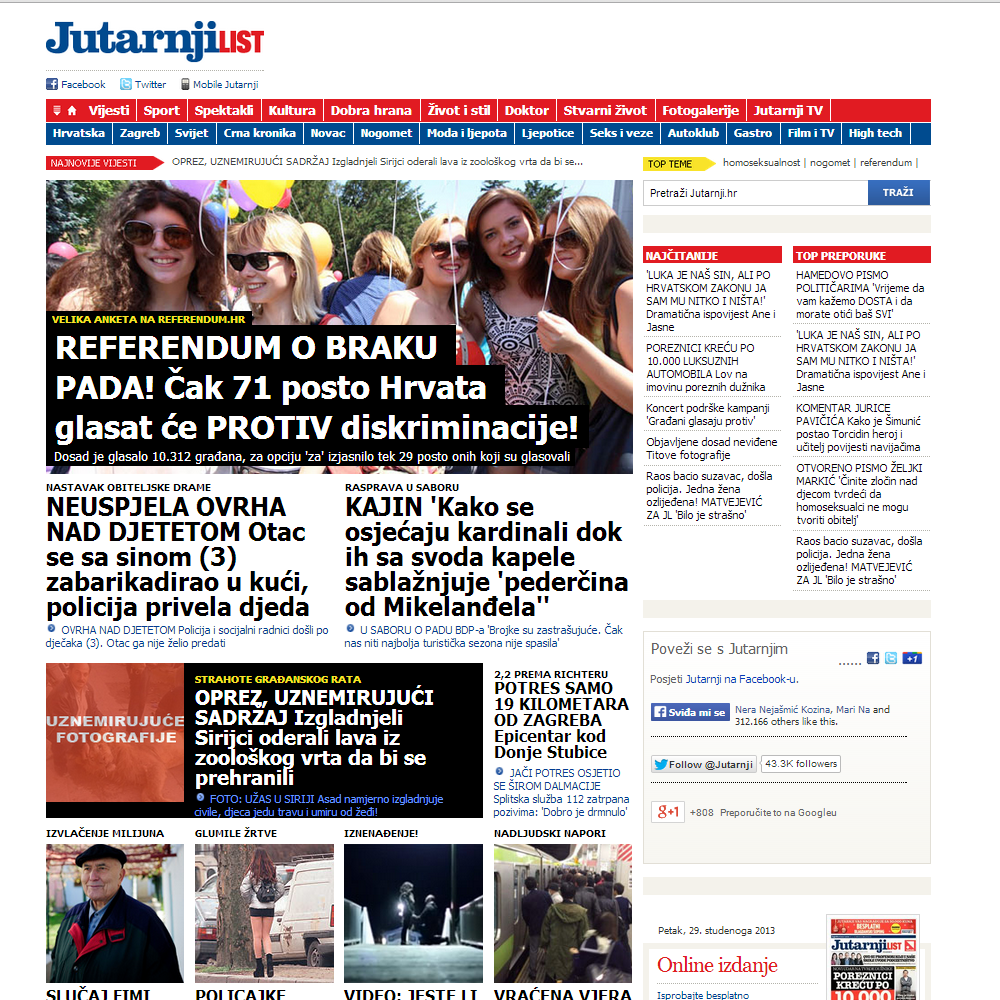}
      \hspace{0.1\textwidth}
      \caption{}
      \label{fig:jutarnji_clanak}
    \end{subfigure}
    \begin{subfigure}[c]{0.33\textwidth}
      \hspace{0.05\textwidth}
      \includegraphics[width=0.9\textwidth]{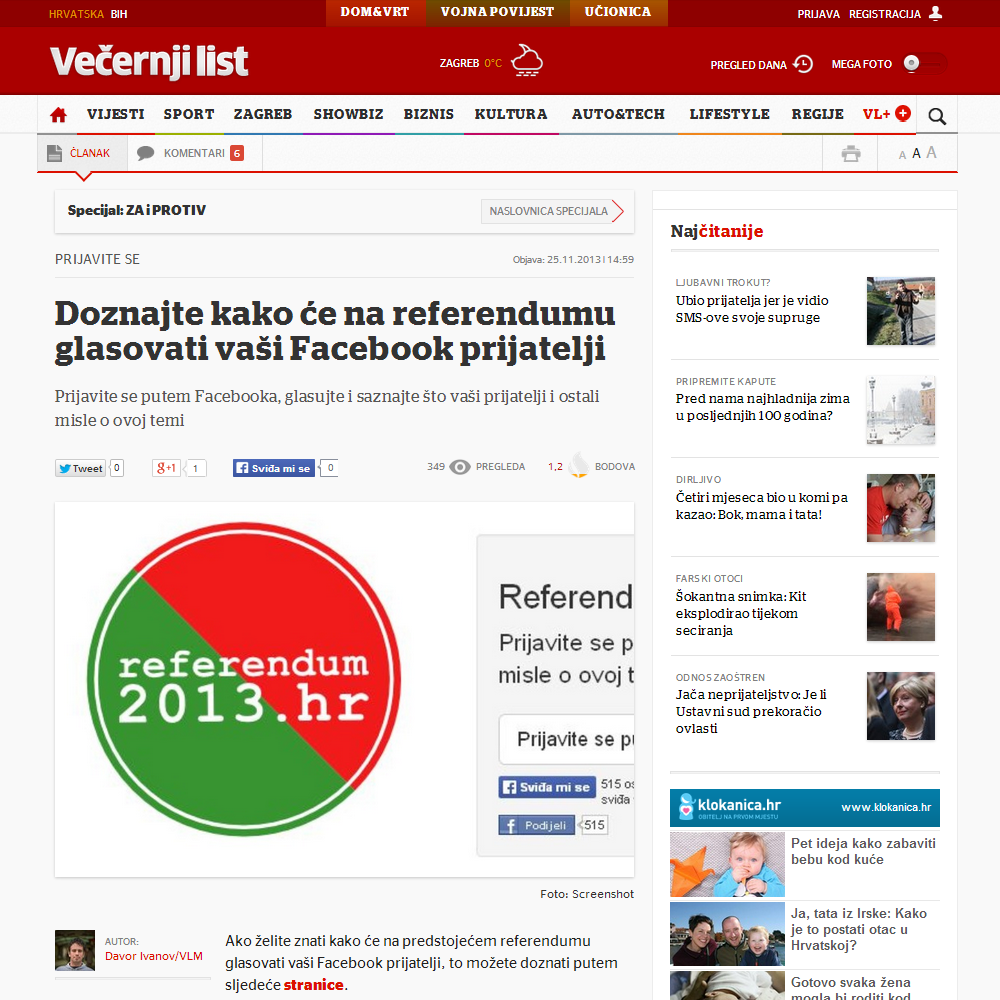}
      \hspace{0.05\textwidth}
      \caption{}
      \label{fig:vecernji_clanak}
    \end{subfigure}
    \begin{subfigure}[c]{0.33\textwidth}
      \hspace{0.2\textwidth}
      \includegraphics[width=0.75\textwidth]{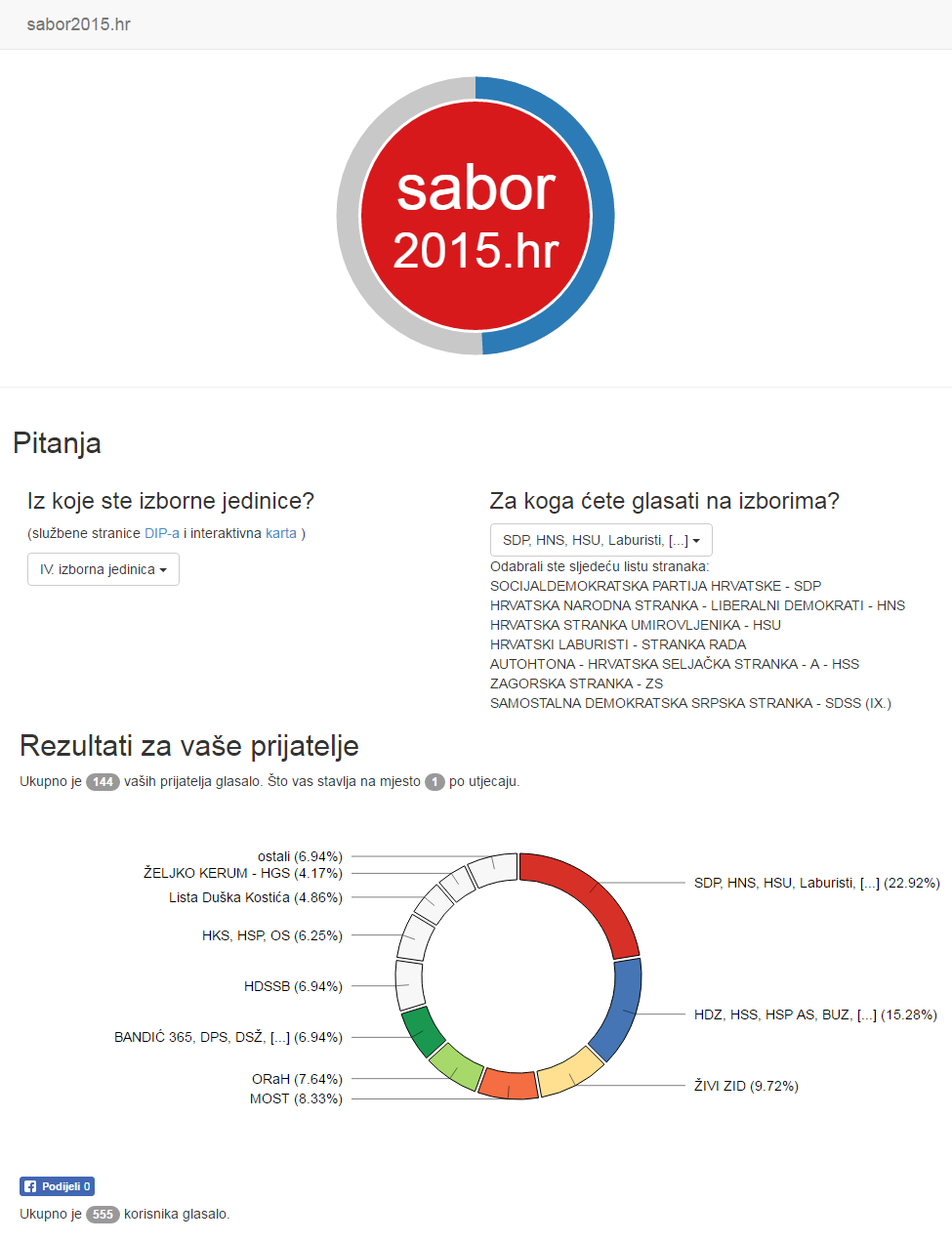}
      \caption{}
      \label{fig:sabor2015_screenshot}
    \end{subfigure}
  \end{subfigure}
\caption{\textbf{News media coverage and a screenshot of survey application.} An example of two online news media articles that reported on the referendum2013.hr application - jutarnji.hr (Figure~\ref{fig:jutarnji_clanak}) and vecernji.hr (Figure~\ref{fig:vecernji_clanak}), with the number of users registered in the 30 minutes window (Figure~\ref{fig:activations_referendum2013}), which is also the window size we used in our inference methodology (more details in the main paper). Sudden peaks in user registrations are closely aligned with the publication of some of the news articles, which indicates possible exogenous influence. Figure~\ref{fig:sabor2015_screenshot} shows the sabor2015.hr online survey application which had a similar interface to the referendum2013.hr application. All three applications allowed Facebook users to cast a vote for the upcoming elections, see votes of their Facebook friends and summary statistics for all registered users.}
\label{fig:activations}
\end{figure}

Once registered, users were given an opportunity to cast their votes in the survey. The web applications did not display any summary statistics before this point so as to not influence user's vote. After voting, user's were shown summary statistics of all users and their Facebook friends separately. To protect privacy of their friends, we did not show statistics on friends if less than a specified number of their friends voted on the survey. Users were able to share the link to the survey application through Facebook. To additionally motivate users to vote and share we displayed a number of users which registered by following their Facebook share, and their rank among the top sharers based on this number. This required of the user to regularly visit the survey application in order to track his status among the sharers and the statistics of his friends, which prolonged user's engagement with the application. Survey applications were typically active only during one week prior to the actual elections (Table~\ref{tab:dataset}). This probably aided in attracting new users as the attention of the news media as well as general public was focused on the possible outcomes of the elections, and our surveys provided a way to satisfy this curiosity.

In addition to the data collected by the Facebook Graph API and our own web server, we also used Google Analytics to manually collect data on the various news coverages which reported on our application. This was especially important for the referendum2013 - our first survey application, as there we did not yet collect any referral links from the users. The exact times and a number of users coming from these web sites (many of which did not register on our application in the end) gives us a qualitative estimate on the magnitude of exogenous influence at these specific times.

During the design and execution of these surveys we tried to follow recommended ethical guidelines for digial social research~\cite{Kosinski2015, Salganik2018}. Every application that uses Graph API has to comply with Facebook's Platform Policy~\footnote{\url{https://developers.facebook.com/policy}} which states conditions under which data can be collected through the API and the responsibilities of the application owners. It specifies the conditions under which user data can be shared to third parties. For example, it explicitly forbids selling the user's data to third parties. It also requires application owners to have a privacy policy which is displayed to the users before they authorize with their Facebook credentials. Our privacy policy which we displayed to the users clearly stated that we will use user's data for research purposes only, and that anonymized data might be made available to the research community in the future. The privacy policies (in Croatian) for the first two survey applications - referendum2013 and sabor2015, are available on a Github open source code repository (Table~\ref{tab:dataset}), along with the source code of the survey application itself. In the main paper we outlined our data sharing policy where we provide access to data upon explicit request and after the interested authors fill in the appropriate agreement. This is becoming an established practice \cite{Vosoughi2018} which aims to satisfy the requirements of reproducible research in cases where free distribution of collected data is restricted by the service provider.

\section{Implementation details}
\label{sec:implementation-details}

Code for statistical inference is vectorized as much as possible by using Scipy's compressed sparse column (CSC) matrix~\footnote{\url{https://docs.scipy.org/doc/scipy/reference/generated/scipy.sparse.csc_matrix.html}} to store adjacency matrix of our friendship network. The CSC format provides efficient addition and multiplication of matrices as well as fast matrix vector products. In addition we sort the matrix elements by the activation times of users in order to exploit the fact that we often slice the friendship matrix into subnetworks of users that activated within a certain time window. This gives performance benefits as slicing a predefined range of a matrix is more efficient than random indexing. Our original likelihood function has the following form:

\begin{equation}
\begin{split}
\mathcal{L}(D;p_{peer},p_{ext},t) = 
&\prod_{i \in \text{activated at } [t - \Delta t, t]} (1-(1-p^{(i)}_{peer}(t))(1-p_{ext}(t))) \\
&\prod_{i \in \text{inactive at } t} (1-p^{(i)}_{peer}(t))(1-p_{ext}(t))
\end{split}
\label{eq:likelihood}
\end{equation}

Note that we could be using more general form of exogenous influence $p^{(i)}_{ext}(t)$ which is user-dependent, but instead we are assuming it is the same for all user at each time step $t$. As multiplication of many small probabilities would soon result in overflow of numerical precision, we exchange multiplication for summation by log-transforming our likelihood. This, however, does not change the value of the maximum likelihood parameters due to the monotonicity of logarithm:

\begin{equation}
\begin{split}
\log \mathcal{L}(D;p_{peer},p_{ext}) = 
&\sum_{i \in \text{activated at } [t - \Delta t, t]} \log (1-(1-p^{(i)}_{peer})(1-p_{ext})) + \\
&\sum_{i \in \text{inactive at } t} \log ((1-p^{(i)}_{peer})(1-p_{ext}))
\end{split}
\label{eq:log-likelihood1}
\end{equation}

For additional numerical stability we also slightly change the way we calculate the log-likelihood. Exogenous influence $p^{(i)}_{ext}$ is assumed to be equal for all users within a specific time frame. The crucial factor here is $p^{(i)}_{peer}$ which is specific for each user and is equal, for SI and EXP models respectively:

\begin{equation}
p^{(i)}_{SI} = 1 - \prod_{j \in N^{(i)} \text{active at } t} (1-p_{SI}) = 1 - (1-p_{SI})^{a_{i}}
\end{equation}

\begin{equation}
p^{(i)}_{EXP}(t) = 1 - \prod_{j \in N^{(i)} \text{active at } t} \left(1-p_{0} e^{-\lambda (t-t_{j})}\right)
\end{equation}

where $N^{(i)}$ is a set of peers of user $i$, $a_{i}$ is the number of activated friends of user $i$, and $p_{SI}$ is a parameter of exogenous influence. However, due to the small probabilities involved we actually calculate equivalent expression using \emph{sum-log-exp} trick:

\begin{equation}
p^{(i)}_{SI} = 1 - \exp \left[ a_{i} \log (1-p_{SI}) \right]
\label{eq:SI_trick}
\end{equation}

\begin{equation}
p^{(i)}_{EXP} = 1 - \exp \left[ \sum_{j \in N^{(i)} \text{active at } t} \log \left(1-p_{0} e^{-\lambda (t-t_{j})}\right) \right]
\label{eq:EXP_trick}
\end{equation}

For this we can use special log1p~\footnote{\url{https://docs.scipy.org/doc/numpy/reference/generated/numpy.log1p.html}} function in Numpy for calculation of $\log(1+x)$, which is more precise when $x$ is small. As for the LOG endogenous influence model which we will introduce in Section~\ref{sec:simulated} (Equation\ref{eq:LOG}), we do not use any additional numerical tricks as in Equations~\ref{eq:SI_trick} and~\ref{eq:EXP_trick}. For optimization we use Scipy's standard \texttt{scipy.optimize.minimize}~\footnote{\url{https://docs.scipy.org/doc/scipy/reference/generated/scipy.optimize.minimize.html}} function with an option \texttt{method=TNC} which implements a truncated Newton algorithm~\cite{Nash2000} which behaves well with the large number of parameters and is Hessian-free so we do not have to provide a gradient of our log-likelihood function. Truncated Newton algorithm uses conjugate gradients to itertively update parameter values, and the inner solver is run for only a limited number of iterations (\emph{truncated}). More details on the scalability analysis is available in Section~\ref{sec:scalability}.

\section{Alternating method for faster convergence of inference}

Algorithm~\ref{alg:alternating_inference} gives the pseudocode of the alternating procedure for inference of endogenous $p_{peer}$ and exogenous $\{p_{ext}\}_{t}$ influence that we use in our experiments. As we already mentioned in the main paper, we are interested in a single set of endogenous influence parameters $p_{peer}$ which are assumed to be constant in time, and a separate set of exogenous influence parameters $\{p_{ext}\}_{t}$ for each time step. Both sets of parameters are assumed to be equal for all users at any given time step. In the first part of the algorithm (steps \ref{alg:step:init1}-\ref{alg:step:init2}) we optimize $p_{peer}$ and $p_{ext}$ for every time window separately, which then serve as initial values for the alternating procedure. Optimization procedure is designated with a generic $MAP$ (Maximum A Posteriori) procedure which takes as arguments the parameters which are held fixed and outputs values of the remaining parameters so that the log-likelihood (Equation~\ref{eq:log-likelihood1}) is maximized. As we already mentioned in Section~\ref{sec:implementation-details}, the actual optimization is performed with a truncated Newton algorithm, although in principle any suitable optimization method could be used. Second part of the algorithm is the actual alternating procedure (steps \ref{alg:step:alternating1}-\ref{alg:step:alternating2}) where we first optimize for a single set of endogenous parameters $p_{peer}$, conditioning on the exogenous parameters $\{p_{ext}\}_{t}$ we obtained for each time window (step \ref{alg:step:map1}). We then optimize exogenous parameters for each window separately $\{p_{ext}\}_{t}$, conditioning on a single set of endogenous parameters $p_{peer}$ we obtained in the previous step (step \ref{alg:step:map2}). We then alternate between the step \ref{alg:step:map1} and \ref{alg:step:map2} until values for $p_{peer}$ and $\{p_{ext}\}_{t}$ converge. The difference between the values for the current and previous iteration are calculated in steps \ref{alg:step:convergence1} and \ref{alg:step:convergence2} and the convergence itself is checked in step \ref{alg:step:alternating1}.

\begin{algorithm}
\caption{Alternating method for joint inference of influence}\label{alg:alternating_inference}
\begin{algorithmic}[1]
\Procedure{AlternatingInference}{$T$,$\epsilon$,$p_{peer}(t)$,$p_{ext}(t)$}
\For{$t\in\{1,...,T\}$} \label{alg:step:init1}
    \State $\{p_{peer}\}_t,\{p_{ext}\}_t \gets$ MAP($p_{peer}(t)$, $p_{ext}(t)$) \Comment{Optimize for every time window.}
\EndFor \label{alg:step:init2}
\While{$\Delta_{peer}^{(i-1)}\geq \epsilon \And \Delta_{ext}^{(i-1)}\geq \epsilon$} \Comment{Until $p_{peer}$ and $\{p_{ext}\}_{t}$ converge.} \label{alg:step:alternating1}
    \State $p_{peer}^{(i)} \gets$ MAP($\{p_{ext}^{(i-1)}\}_{t}$) \Comment{Fix $\{p_{ext}^{(i-1)}\}_{t}$ and optimize for single $p_{peer}^{(i)}$.} \label{alg:step:map1}
    \State $\{p_{ext}\}^{(i)} \gets$ MAP($p_{peer}^{(i)}$) \Comment{Fix $p_{peer}^{(i)}$ and optimize $\{p_{ext}^{(i)}\}_{t}$ for every window.} \label{alg:step:map2}
    \State $\Delta_{peer}^{(i)}\gets p_{peer}^{(i)}-p_{peer}^{(i-1)}$ \label{alg:step:convergence1}
    \State $\Delta_{ext}^{(i)}\gets \sum_{t=1}^{T} (p_{ext}^{(i)}(t)-p_{ext}^{(i-1)}(t))$ \label{alg:step:convergence2}
    \State $i\gets i+1$
\EndWhile \label{alg:step:alternating2}
\State \textbf{return} $p_{peer}^{(i)},\{p_{ext}^{(i)}\}_{t}$ \Comment{The parameters of endogenous and exogenous influence.}
\EndProcedure
\end{algorithmic}
\end{algorithm}

Even when using this alternating method our inference could still fail to converge, especially in the cases of the two-parameter exponential decay model $p^{(i)}_{EXP}(t) = p_{0} \exp[\lambda(t-t')]$, where parameters are $(p_{0},\lambda)$, and logistic threshold model $p^{(i)}_{LOG}(t) = 1 / (1 + e^{-k(a_{i}(t)-a_{0})})$, where parameters are $(k,a_{0})$. In that case we can choose several reasonable values for a single parameter, for example $\lambda$, and optimize log-likelihood separately for each of these cases where value of this parameter is fixed. We can then choose among these the parameter value which yields the best log-likelihood. In this way we are effectively optimizing multiple 1-parameter models instead of a single 2-parameter model. This could even be done in parallel in order to gain speed benefits. The same method was used in~\cite{Myers2012} paper where they reduced 2-parameter model for an \emph{exposure curve}, which describes individual's susceptibility to endogenous influence, into a 1-parameter model.

\section{Calculation of individual and collective influence}

Our assumption regarding \emph{individual influence} is that each user is, to some extent, responsible for the endogenous activation of all of his peers. To illustrate this we demonstrate how to calculate influence in a simple example shown on the Figure bellow where we have a social network of five users ($u_1$,$u_2$,$u_3$,$u_4$ and $u_5$) that activated due to endogenous or exogenous influence, each at a specific time $t_i$. Arrows indicate potential for endogenous influence - connections from users that activated before are pointed towards users that activated after. We start with the input data - adjacency matrix $A$ which encodes the friendship connections between users, an array of activation times for all users and an array which encodes whether user activated due to endogenous (1) or exogenous (0) influence:

\noindent
\begin{minipage}{.5\linewidth}
\begin{align*}
    A &=
    \begin{bmatrix}
      0 & 1 & 0 & 1 & 0 \\
      1 & 0 & 1 & 0 & 1 \\
      0 & 1 & 0 & 0 & 0 \\
      1 & 0 & 0 & 0 & 1 \\
      0 & 1 & 0 & 1 & 0
    \end{bmatrix}
    \\
    \text{activation times} &= 
    \begin{bmatrix}
      1 & 2 & 3 & 4 & 5
    \end{bmatrix}
    \\
    \text{endogenous activation} &= 
    \begin{bmatrix}
      0 & 0 & 1 & 1 & 1
    \end{bmatrix}
\end{align*}
\end{minipage}%
\begin{minipage}{.5\linewidth}
\centering
\includegraphics[width=0.6\linewidth]{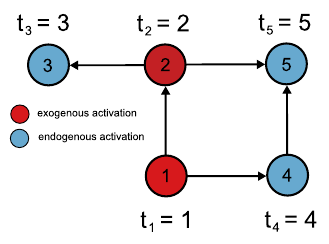}
\end{minipage}

Out of these, the endogenous activation is actually not available in empirical data, and has to be inferred. For the purpose of our example we assume we somehow estimated it, either from raw data or using an inference methodology such as ours. Let us calculate several measure which will help us in our calculation - a number of peers which were active \emph{before} each user's activation and a number of peers which were active \emph{after} each user's activation:

\begin{align*}
\text{number of active peers \emph{at} activation} &= \begin{bmatrix}
       0 & 1 & 1 & 1 & 2
     \end{bmatrix}
\\
\text{number of active peers \emph{after} activation} &= \begin{bmatrix}
       2 & 2 & 0 & 1 & 0
     \end{bmatrix}
\end{align*}

Users activated sequentially. For the user $u_1$, there are two of his peers that activated after, but only one due to endogenous influence (user $u_4$). There are no other peers of user $u_4$ that activated before him, so user one gets full credit for his endogenous activation, and his individual influence is 1.0. User $u_2$ has two peers that activated after him - users $u_3$ and $u_5$, and both activated due to endogenous influence. User $u_3$ has no other peers that activated before him, so user $u_2$ gets full credit and his individual influence 1.0. User $u_5$ has one more peer who activated before him (user $u_4$) so user $u_2$ gets only half of the credit (individual influence 0.5), making a total of 1.5. The final individual influence $I$ for all users is:

\begin{equation*}
\text{I} = \begin{bmatrix}
       1.0 & 1.5 & 0.0 & 0.5 & 0.0
     \end{bmatrix}
\end{equation*}

In the above example we have used the simplest expression for the calculation of individual influence $I^{(i)}$ of user $i$ which we already introduced in the main paper:

\begin{equation}
I^{(i)} = \sum_{j \in N^{(i)}} \frac{I^{(i \to j)}}{\sum_{m \in N^{(j)}} I^{(m \to j)}} p^{(j)}_{peer}
\label{eq:individual_influence}
\end{equation}

In this case, the quantity $I^{(i \to j)}$ is simply $1$, meaning that each user has equal influence on all of his peers which activated after him, regardless of how far away in time this actually is. A more realistic formulation is to make $I^{(i \to j)}$ dependent on time, so that the users who activated closes in time to their peers are credited with more influence. One way of introducing time dependency is with a simple exponential decaying influence, in which case $I^{(i \to j)} = e^{-\lambda(t_{j}-t_{i})}$. The new equation for the influence $I^{(i)}$ of user $i$ is now:

\begin{equation}
I_{\text{EXP}}^{(i)} = \sum_{j \in N^{(i)}} \frac{e^{-\lambda(t_{j}-t_{i})}}{\sum_{m \in N^{(j)}} e^{-\lambda(t_{j}-t_{m})}} p^{(j)}_{peer}
\label{eq:individual_influence_new}
\end{equation}

We should note that the choice of how we calculate individual influence, either through $I^{(i)}$, $I_{\text{EXP}}^{(i)}$ or some other formulation, is independent on the choice of particular endogenous influence model $p_{peer}$. In our experiments on empirical data we are using formulation of individual influence from Equation~\ref{eq:individual_influence} for both the SI and EXP endogenous influence models.

\section{Scalability of inference}
\label{sec:scalability}

\begin{figure}[ht]
\centering
\includegraphics[width=0.4\linewidth]{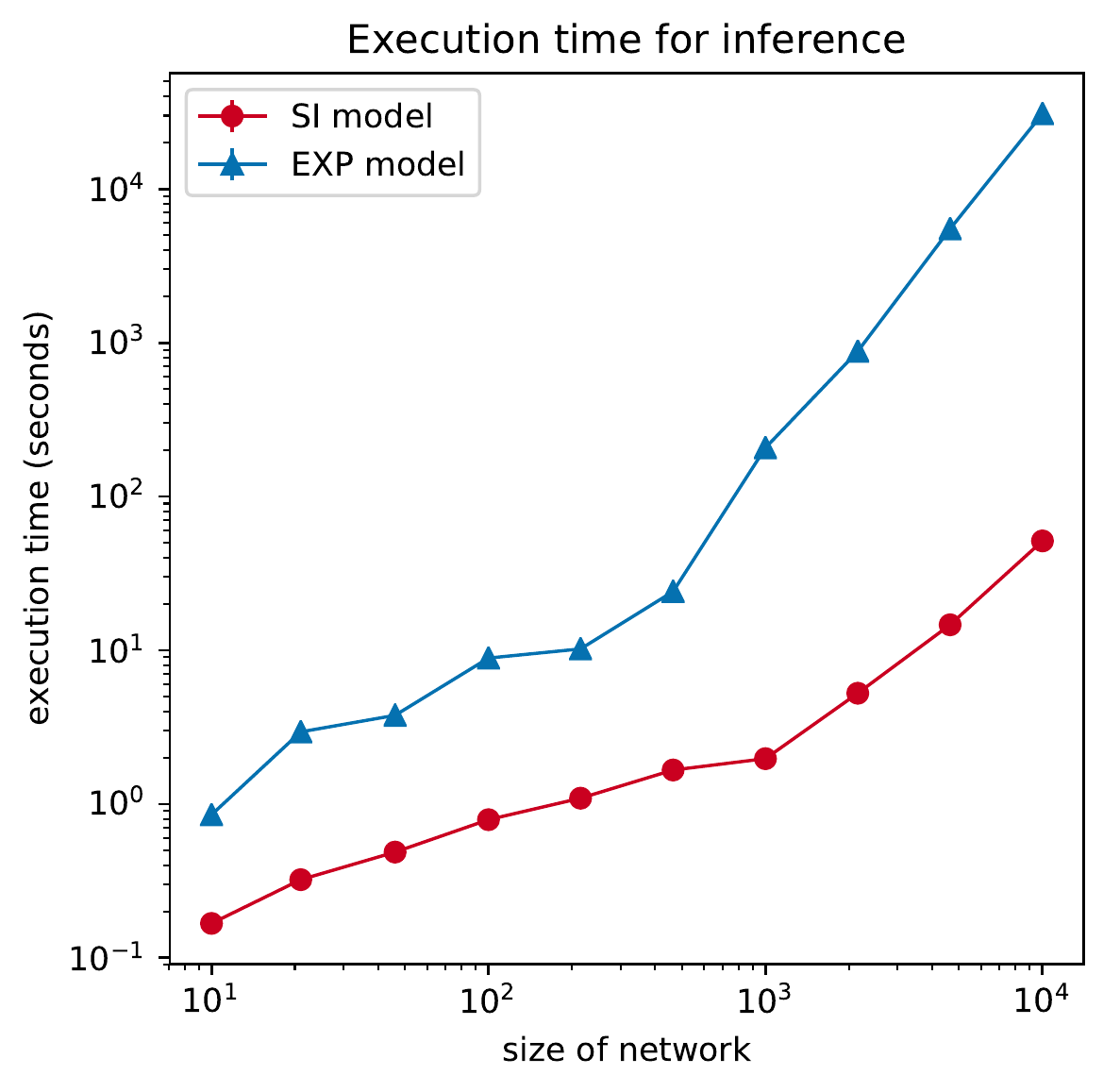}
\caption{\textbf{Scalability analysis.} Scalability analysis is based on inference experiments performed on simulated activation cascades following SI and EXP models. Execution times rise are almost linear with respect to the size of the networks.}
\label{fig:scalability}
\end{figure}

We test the scalability of our method by inference on simulated cascades and increasingly larger networks. We construct the networks with NetworkX's \texttt{powerlaw\_cluster\_graph}~\footnote{\url{https://networkx.github.io/documentation/networkx-1.10/reference/generated/networkx.generators.random_graphs.powerlaw_cluster_graph.html}} function which implements Holme and Kim algorithm for generating networks with powerlaw degree distribution and desired average clustering. At each step of the algorithm we add a node with 3 new edges (\texttt{m=3}) and set the clustering probability to 0.1 (\texttt{p=0.1}). We explore graphs of sizes ranging from $10$ to $1000$ (Figure~\ref{fig:scalability}). Execution times are almost linear with respect to the size of the networks on which inference is being done. The inference was run on a 64-bit Intel i5-2500 CPU 3.3 GHz and 8 GB of RAM, Python 3.6.1. as a part of Anaconda distribution.

\section{Correction for the observer bias}
\label{sec:observer_bias}

Due to the particular way we collected our survey data - our friendship network consists only of users who activated eventually in one week time period during which the survey application was active, we tend to overestimate of exogenous influence as we approach the end of the observation period. The reason for this \emph{observer bias} is because our set of inactive users is getting smaller, but the rate of registrations is independent of this and so we overestimate exogenous influence. We correct for this by artificially extending our friendship network of inactive users by a certain fraction $\alpha$ of the total friendship through factor $c(t)$:

\begin{equation}
c(t) = 1 + \alpha \frac{N_{\text{all}}}{N_{\text{inactive}}(t)}
\label{eq:correction}
\end{equation}

, where $N_{all}$ is the number of all users in the social network, and $N_{inactive}(t)$ the number of all yet users inactive users at time $t$. And including it into our log-likelihood function (Equation~\ref{eq:log-likelihood1}) to modify the part with inactive users:

\begin{equation}
\begin{split}
\log \mathcal{L}(D;p_{peer},p_{ext},t) = 
&\sum_{i \in \text{activated at } [t - \Delta t, t]} \log (1-(1-p^{(i)}_{peer}(t))(1-p_{ext}(t))) + \\
&c(t) \sum_{i \in \text{inactive at } t} \log ((1-p^{(i)}_{peer}(t))(1-p_{ext}(t)))
\end{split}
\label{eq:log-likelihood2}
\end{equation}

In the expression for $c(t)$, $\alpha$ is the size of the virtual friendship network with which we want to expand our observed network, expressed as a fraction of the observed friendship network. In case of $\alpha=0$ we are not making any expansion at all, and we can expect this observer bias effect to exaggerate exogenous influence as the size of the activation cascade approaches the size of the observed friendship network. 

\begin{figure}[ht]
  \centering
  \includegraphics[width=1.0\linewidth]{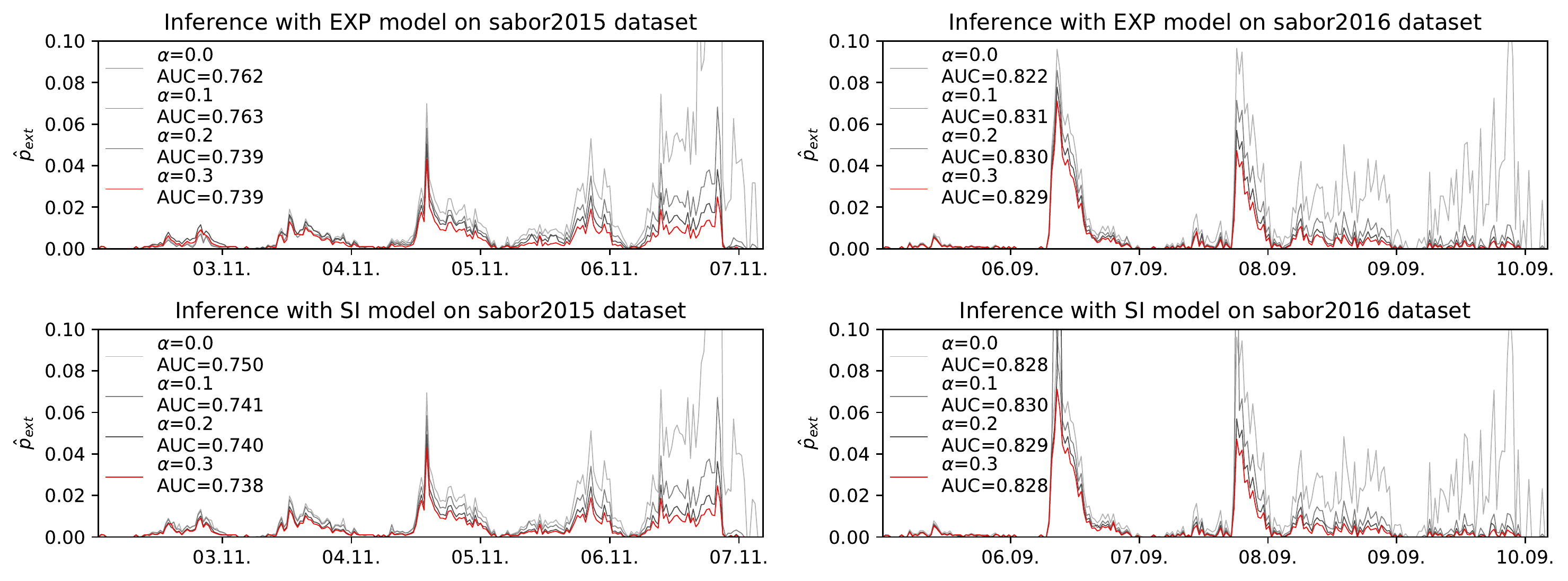}
  \caption{\textbf{Correction for the observer bias.} Inference of endogenous and exogenous influence on sabor2015 and sabor2016 empirical datasets with various values $\alpha$ of the observer bias correction factor and SI and EXP as the assumed endogenous influence models. We tried four different values of $\alpha$ ranging from $0.0$ (no correction) to
  $0.3$ which corresponds to the increase of the virtual friendship network of users by $0-30\%$. Observer bias arises due to the fact that we only collect friendship relations between users that registered on our application until the end of the collection period - the pool of inactivate users is much larger than what we actually observe, and this influences our estimates of exogenous influence. The effect of observer bias is the exogenous influence artificially increases as we approach the end of the observation period. However, because there is less and less registered users as we approach the end of the observation period, the correction by itself does not significantly increase the predictive power of our inference methodology as judged by the AUC scores.}
  \label{fig:correction_observer_bias}
\end{figure}

Figure~\ref{fig:correction_observer_bias} shows the effect of the correction for the observer bias on the empirical datasets we collected. We only perform the experiments on the sabor2015 and sabor2016 datasets because for them we have the information on the referral links from which users visited our survey application, which effectively gives us the means to evaluate our estimates of endogenous and exogenous influence for each individual user. We see that corrections with $\alpha=0.1$ and higher successfully reduce the artificially high values of exogenous influence near the end of the observation period. However, the effect of the correction on the predictive power of our model is not large, as judged by the AUC scores. This is mainly due to the fact that the activations fall of near the end of the observation period and so the effect of correcting them is almost negligible.

\section{Experiments on simulated data (extended)}
\label{sec:simulated}

\begin{figure}[ht]
  \centering
  \begin{subfigure}[c]{0.49\textwidth}
    \includegraphics[width=1.0\linewidth]{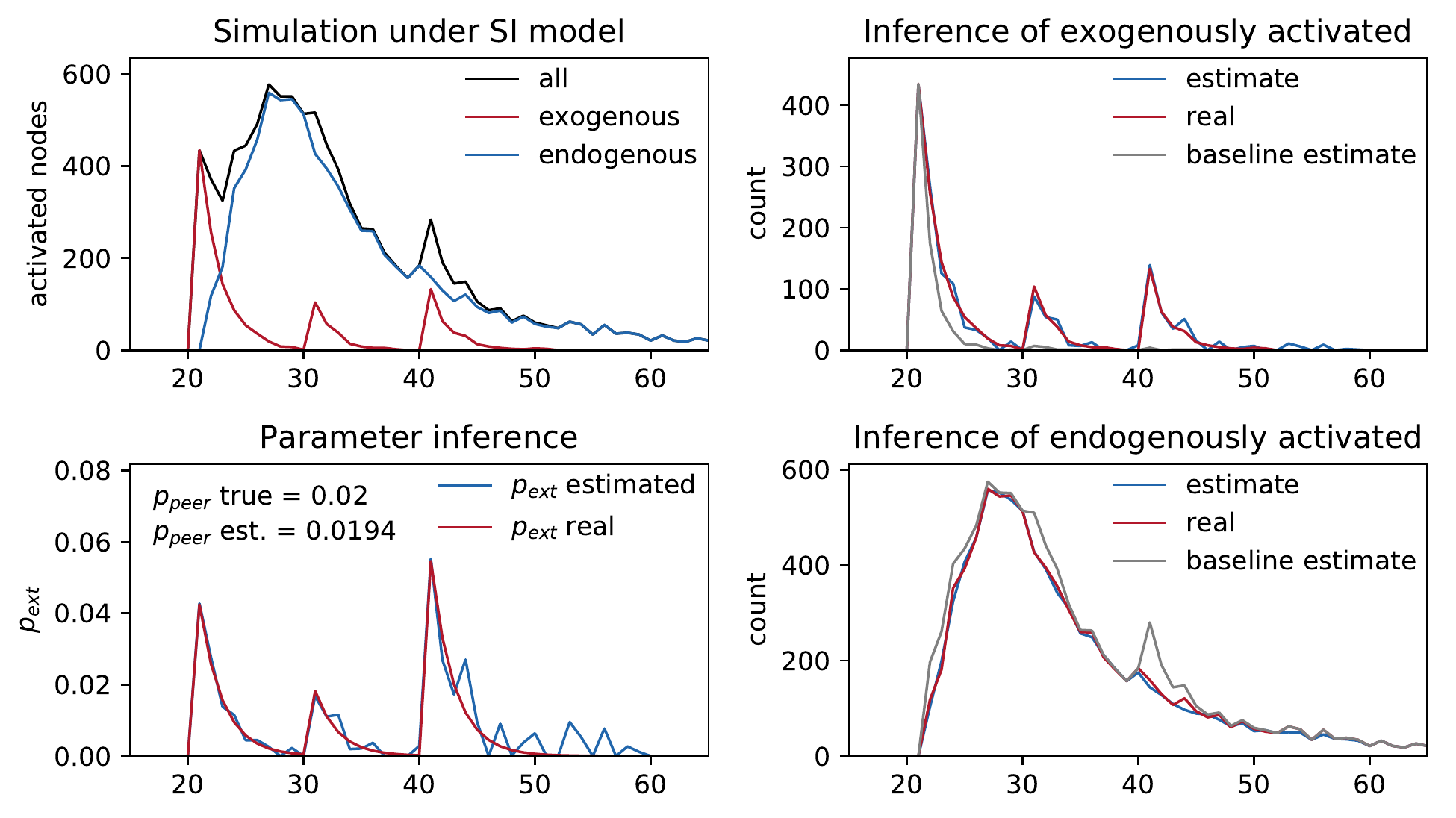}
    \includegraphics[width=1.0\linewidth]{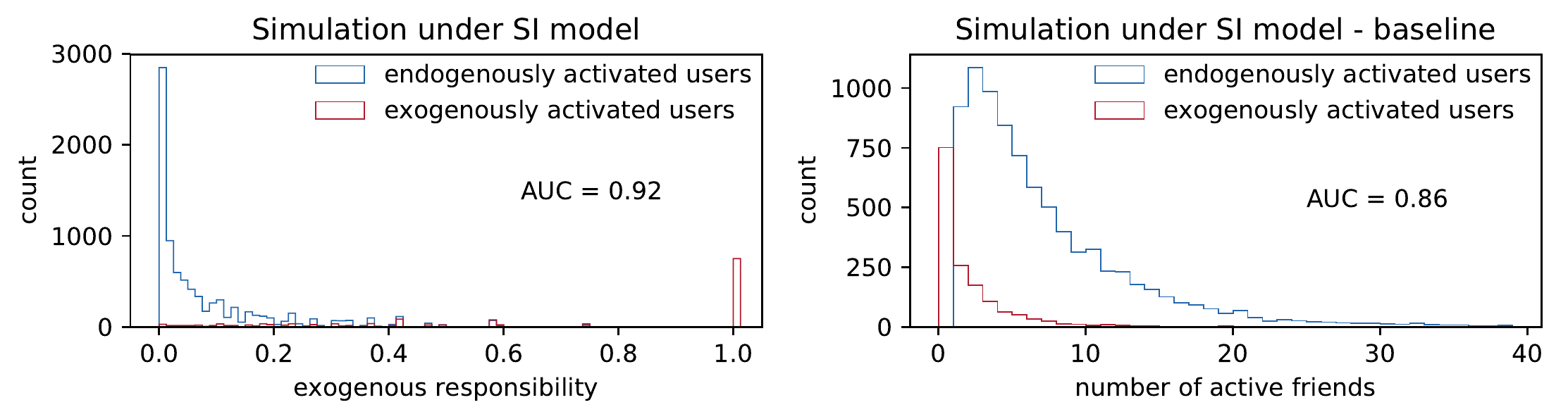}
    \caption{}
    \label{fig:simulation_SI}
  \end{subfigure}
  \begin{subfigure}[c]{0.49\textwidth}
    \includegraphics[width=1.0\linewidth]{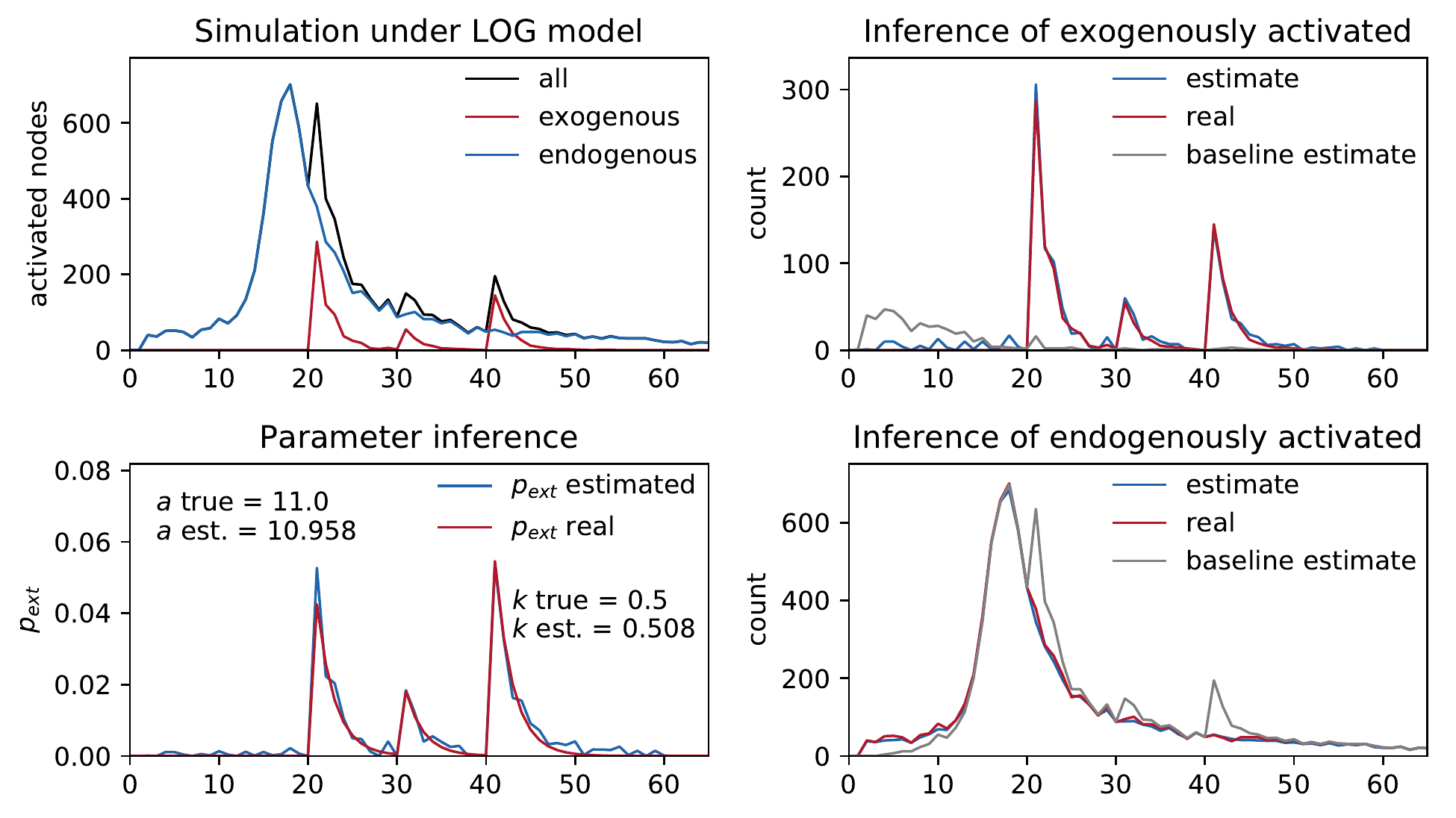}
    \includegraphics[width=1.0\linewidth]{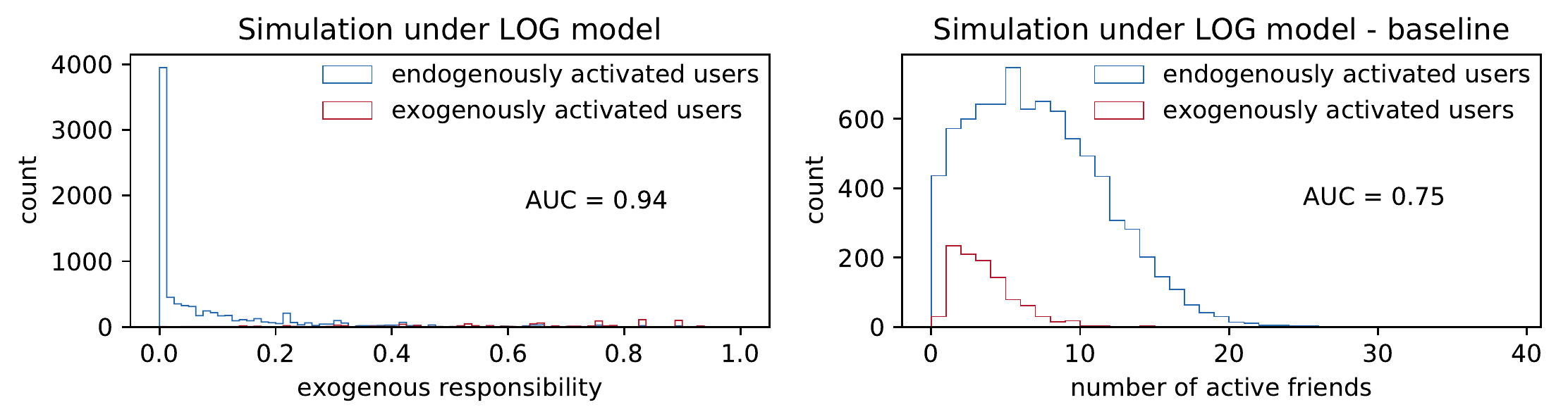}    
    \caption{}
    \label{fig:simulation_LOG}
  \end{subfigure}
  \caption{\textbf{Inference on simulated activation cascades on SI and LOG models.} We performed additional inference experiments using SI (Figure~\ref{fig:simulation_SI}) and LOG (Figure~\ref{fig:simulation_LOG}) models for endogenous influence, which are representative examples of \emph{simple} (SI) and \emph{complex} (LOG) contagion models. We are able to infer correct values of endogenous and exogenous parameters as well as the total estimates for uses activated due to the one or the other influence. Same as in experiments in the main paper, we modeled the exogenous influence so that it resembles typical situation when external news source activates some of the users - with spiked exponentially decaying shape.}
  \label{fig:simulation_extended}
\end{figure}

In this section we show the results of the inference on simulated activation cascades following the susceptible-infected (SI) model (Figure~\ref{fig:simulation_SI}). We also introduce one additional model - a logistic threshold (LOG) model, where we do not assume independence of activation cascades in calculation of endogenous activation probability but rather derive it using the number of already active peers $N^{(i)}$ of user $i$. LOG model is an example of \emph{complex contagion} model and we define it in the following way:

\begin{equation}
p^{(i)}_{LOG}(t) = \frac{1}{1 + e^{-k(a_{i}(t)-a_{0})}}
\label{eq:LOG}
\end{equation}
Where $a_{i}(t)$ designates how many of user $i$'s peers are active at time $t$, and $a_{0}$ and $k$ are parameters of endogenous influence which define the shape of sigmoidal activation function, with $a_{0}$ being a number of active friends you need for the probability of activation to reach $0.5$.

Similar as the experiment on the exponential decay (EXP) model, our methodology is able to correctly infer the parameters of both SI (Figure~\ref{fig:simulation_SI}) and LOG (Figure~\ref{fig:simulation_LOG}) endogenous influence models. We also estimate an absolute number of users activated due to endogenous or exogenous influence, as well as characterize the activation of each user as being driven dominantly by one or the other influence. In theory, any microscopic model of endogenous influence can be used which can be efficiently computed given information on user's friendship connections and the activation state of his peers. This includes \emph{simple contagion} model where activation probability of users is independent of the rest of the system (in our case these are SI and EXP models) as well as \emph{complex contagion models} where probability of activation is dependent on the state of the user's neighborhood of peers (in our case this is LOG model).

\begin{figure}[ht]
  \centering
  \includegraphics[width=\textwidth]{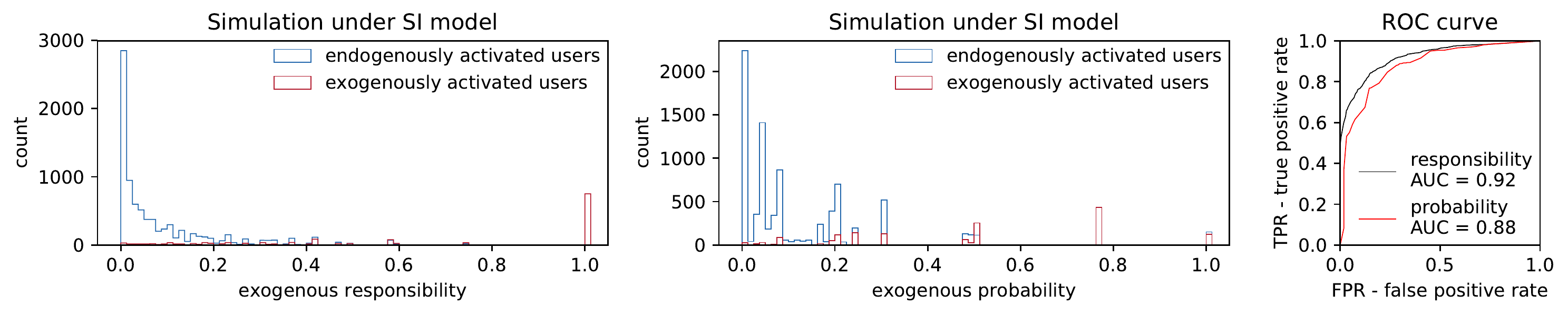}
  \includegraphics[width=\textwidth]{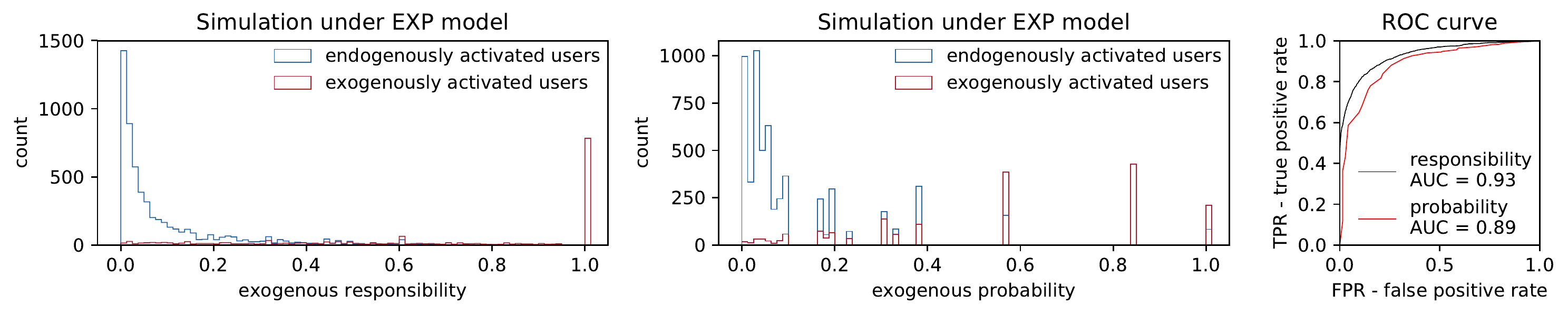}
  \includegraphics[width=\textwidth]{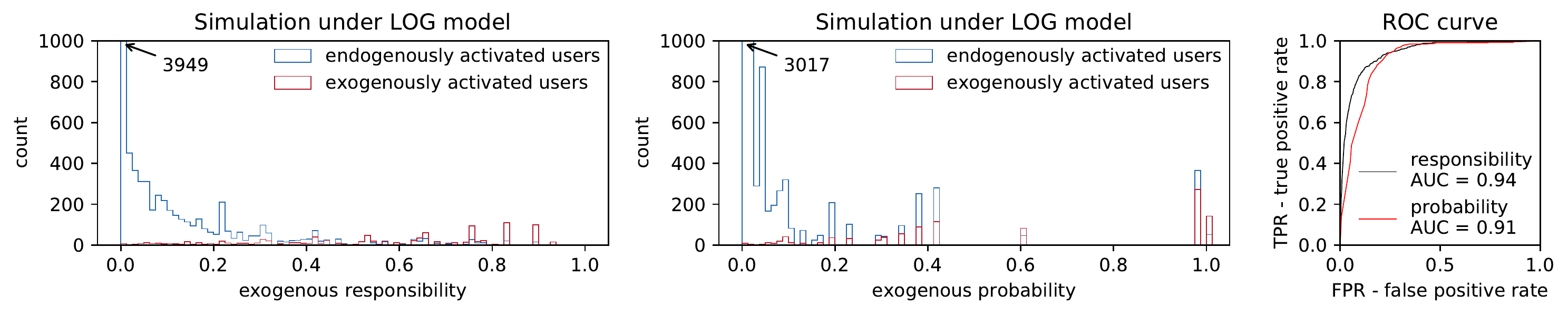}
  \caption{\textbf{Inference on simulated activation cascades using exogenous activation probability.} Histograms of of exogenous activation probability $p^{(i)}_{\text{ext}}$ (middle figures) and exogenous responsibility (Equation \ref{eq:responsibility}, left figures) obtained from simulated activation cascades. Both measures can be used for quantifying the magnitude of exogenous influence. Experiments are performed on  three types of endogenous influence models - SI (top figures), EXP (middle figures) and LOG (bottom figures). Estimates obtained with $p^{(i)}_{\text{ext}}$ alone are coarser due to the fact that the ground truth exogenous influence assumes just a few distinct values. Also, the AUC scores are worse than when using the exogenous responsibility measure. }
  \label{fig:exogenous_activation_probability}
\end{figure}

We also perform simulated experiments where we use an alternative measure for estimating exogenous influence - instead of external responsibility we use normalized exogenous activation probability $p^{(i)}_{\text{ext}}$ directly. Figure \ref{fig:exogenous_activation_probability} shows the results of using exogenous activation probability instead of exogenous responsibility for estimating magnitude of exogenous influence. The parameters of the endogenous influence and the shape of the exogenous influence are kept the same as in the previous simulated experiment in Figure~\ref{fig:simulation_extended}. We see that using the exogenous activation ability alone makes coarser estimates of the influence, reflecting the fact that the ground truth exogenous influence assumes just a few distinct values during the several time windows. Also, the AUC scores are worse than when using the exogenous responsibility measure which incorporates endogenous influence as well.

\begin{figure}[ht]
  \centering
  \includegraphics[width=1.0\linewidth]{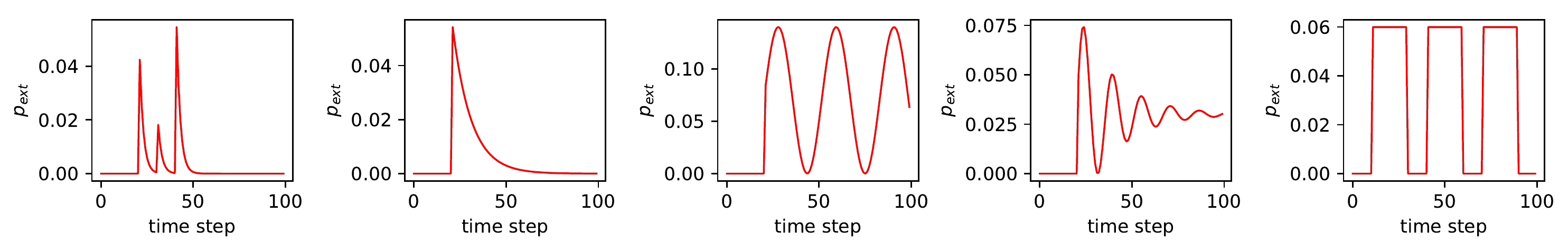}
  \caption{\textbf{Different shapes of exogenous influence.} We perform extended experiments on simulated activation cascades using different shapes of exogenous influence. Exponentially decaying peaks in activity (first two graphs) are typical in realistic cases and are usually caused by various news media announcements (Figure 1c of the main paper). Other shapes are less realistic but we use them to show the robustness of our inference methodology.}
  \label{fig:event_profiles}
\end{figure}

In our main paper we decided to use spike-shaped exogenous influence in our synthetic experiments. This shape closely resembles what would we expect in real cases when the sudden surge in user registration is caused by a media news coverage coming from an external source which then dissipates exponentially in time. Exponentially-decaying spikes of user activity are also observed in Google search queries related to sudden catastrophic events \cite{Crane2008}. However, our inference methodology can easily handle exogenous influence with other shapes - for example constant, exponentially decaying and sinusoidal (Figure~\ref{fig:event_profiles}). In theory, as our method is nonparametric and is calculated at each time step, we could perform a correct inference even in cases when exogenous influence is functionally dependent on some of the dynamically changing properties of the activation cascade, for example a number of activated users.

\begin{figure}[h!t]
  \centering
  \begin{subfigure}[c]{0.49\textwidth}
    \includegraphics[width=1.0\linewidth]{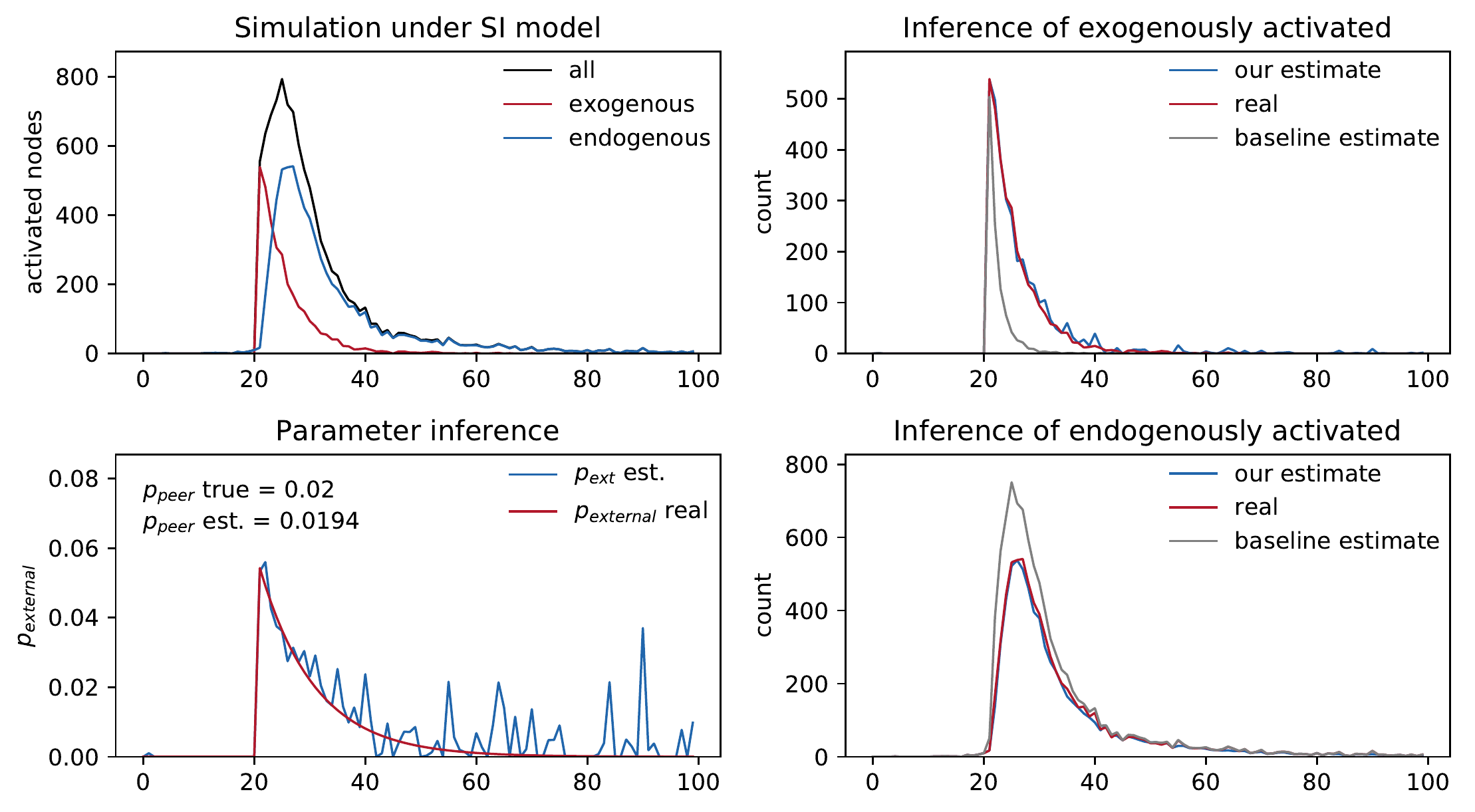}
    \caption{}
    \label{fig:simulation_event_profile1}
  \end{subfigure}
  \begin{subfigure}[c]{0.49\textwidth}
    \includegraphics[width=1.0\linewidth]{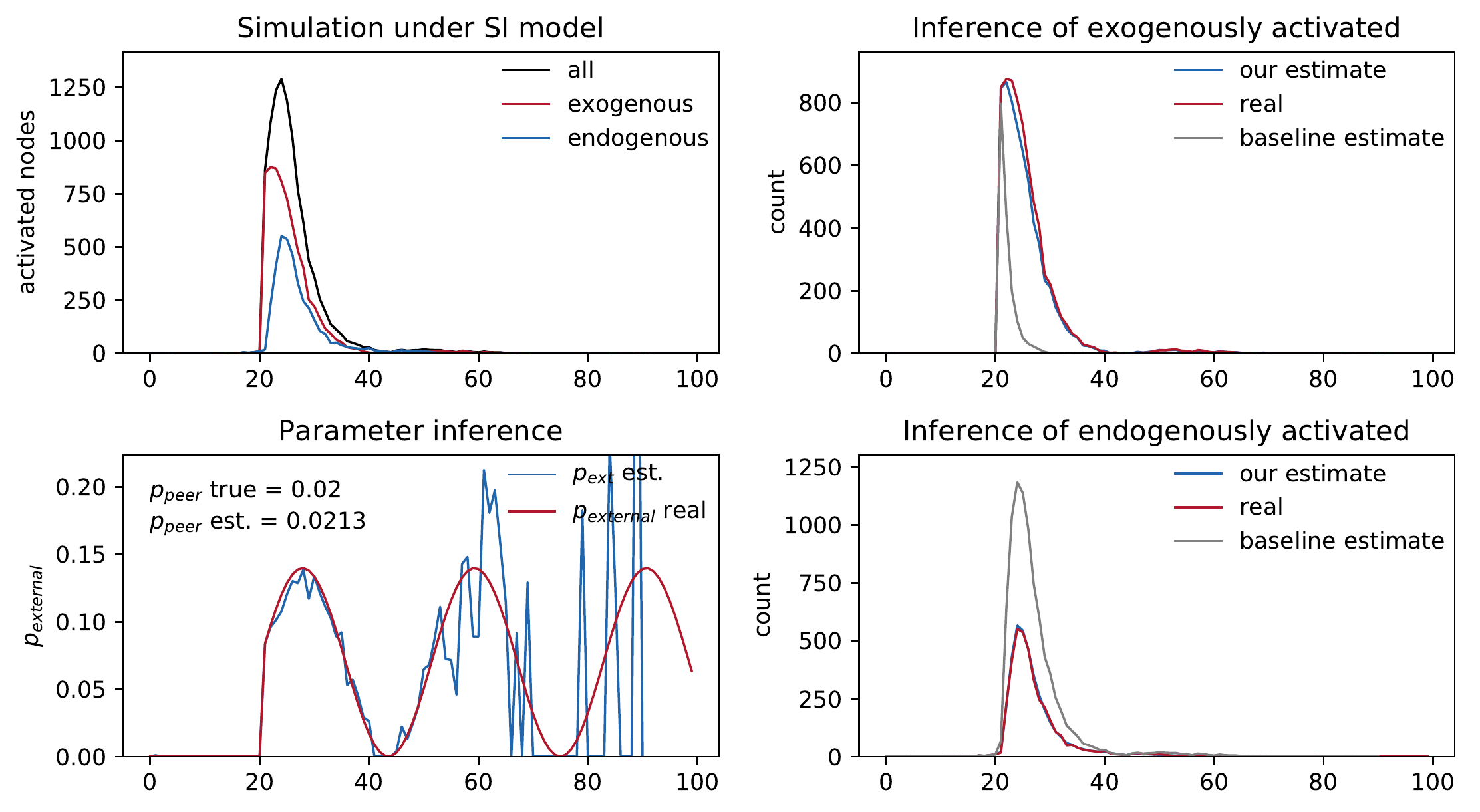}
    \caption{}
    \label{fig:simulation_event_profile2}
  \end{subfigure}
  \begin{subfigure}[c]{0.49\textwidth}
    \includegraphics[width=1.0\linewidth]{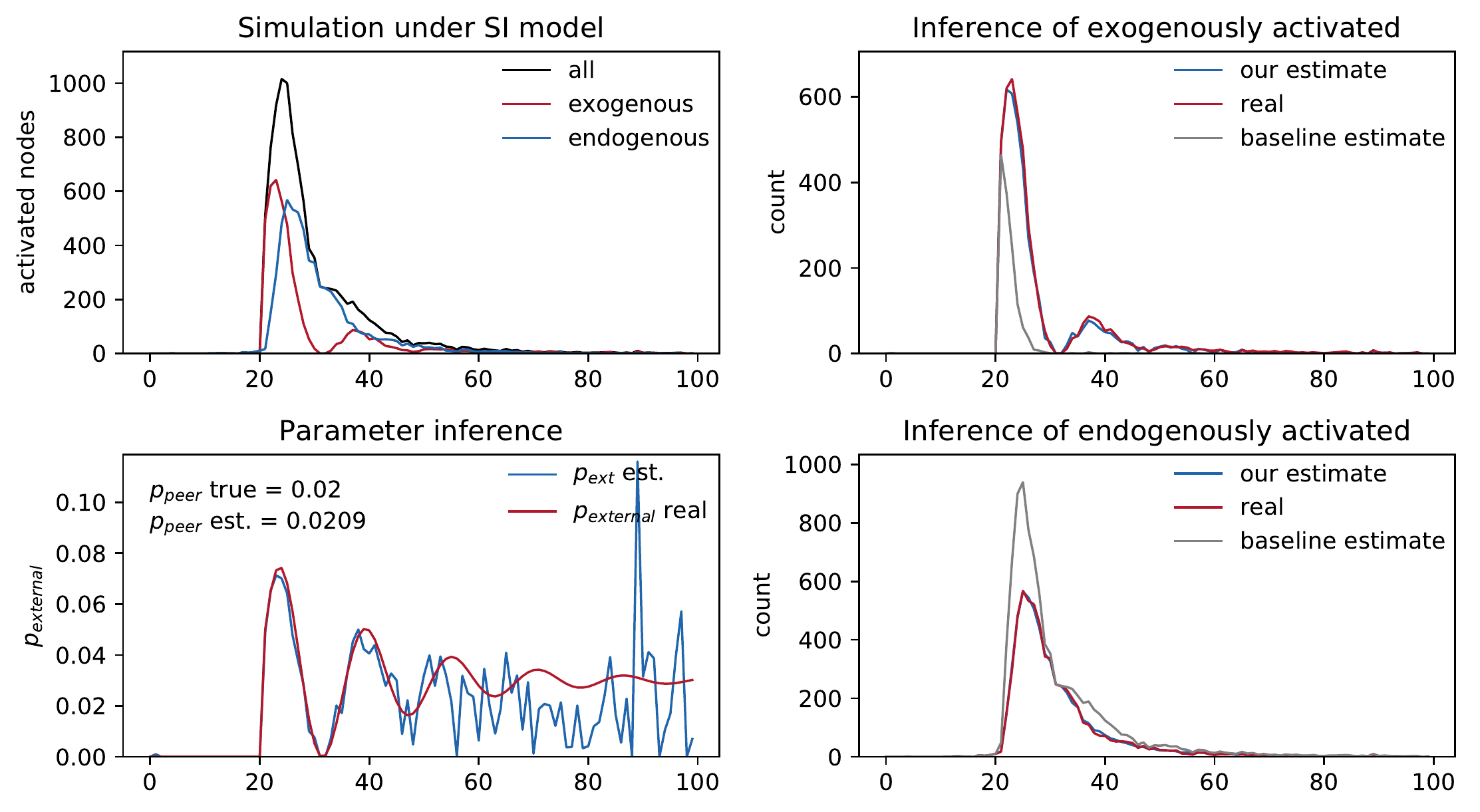}
    \caption{}
    \label{fig:simulation_event_profile3}
  \end{subfigure}
  \begin{subfigure}[c]{0.49\textwidth}
    \includegraphics[width=1.0\linewidth]{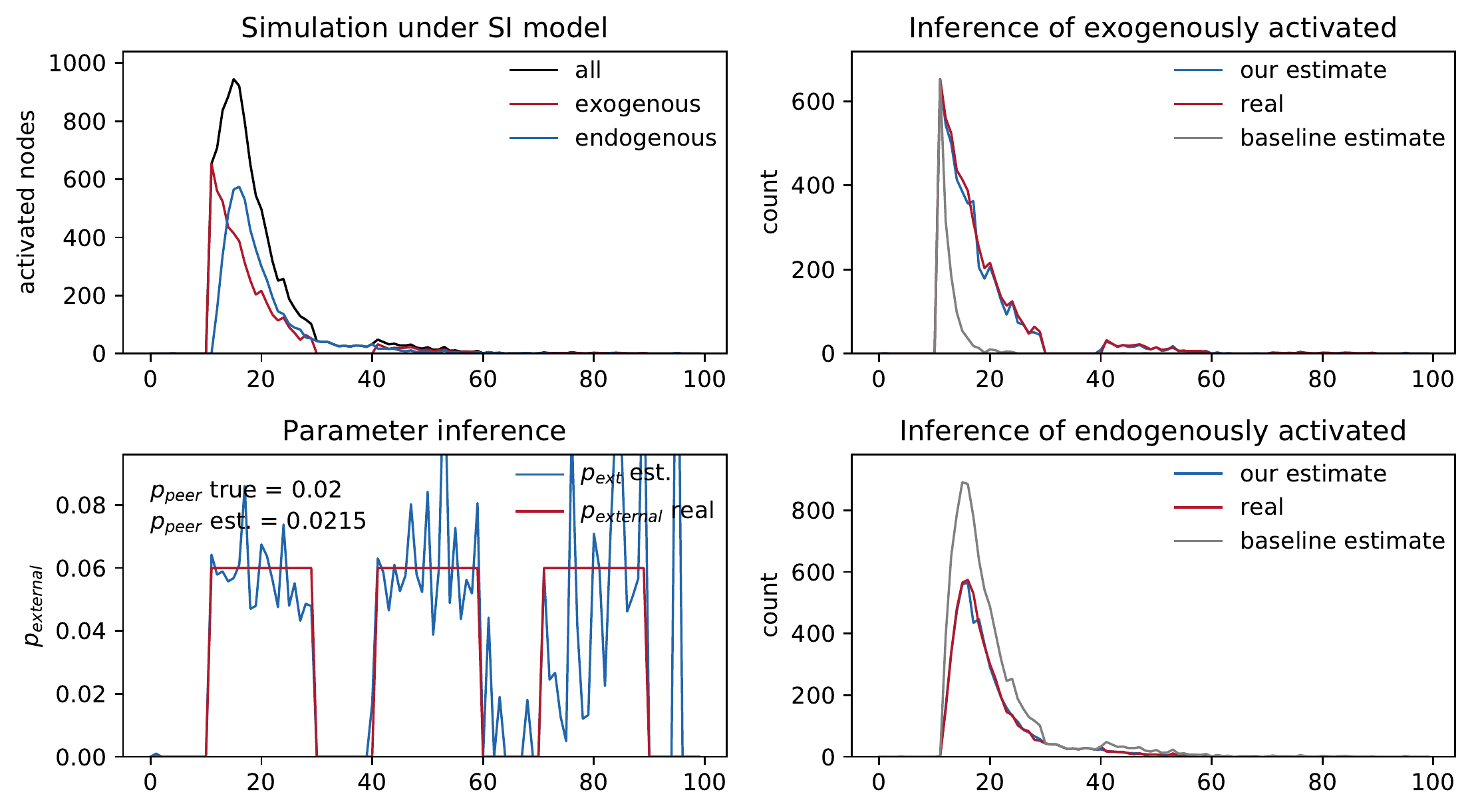}
    \caption{}
    \label{fig:simulation_event_profile4}
  \end{subfigure}
  \caption{\textbf{Simulated activation cascades with different shapes of exogenous influence.} Our inference methodology is robust to the shape of exogenous influence and is able to infer parameter values of both endogenous and exogenous influence in all cases. In the simulated activation cascades above we use the four exogenous influence curves from Figure~\ref{fig:event_profiles} (without the first curve which we use in all other experiments on simulated data), and SI as the model of endogenous influence.}
  \label{fig:different_shapes}
\end{figure}

\begin{figure}[ht]
  \centering
  \begin{subfigure}[c]{0.33\textwidth}
    \includegraphics[width=1.0\linewidth]{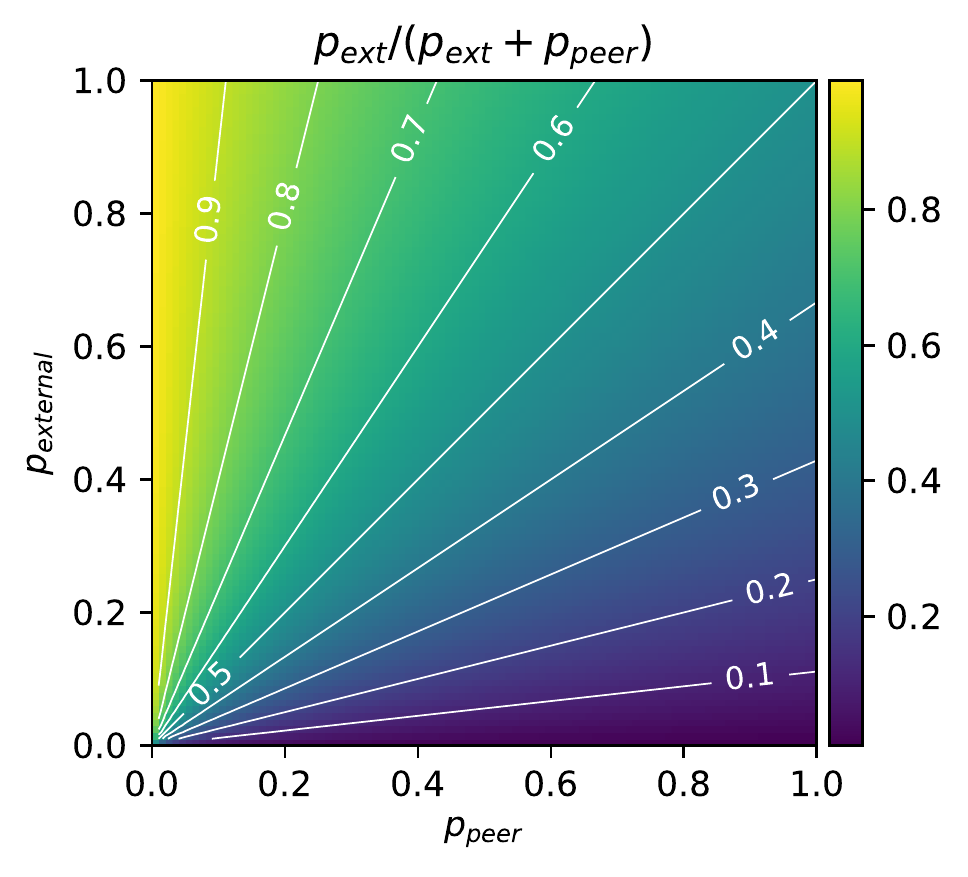}
    \caption{}
    \label{fig:responsibility}
  \end{subfigure}
  \begin{subfigure}[c]{0.33\textwidth}
    \includegraphics[width=1.0\linewidth]{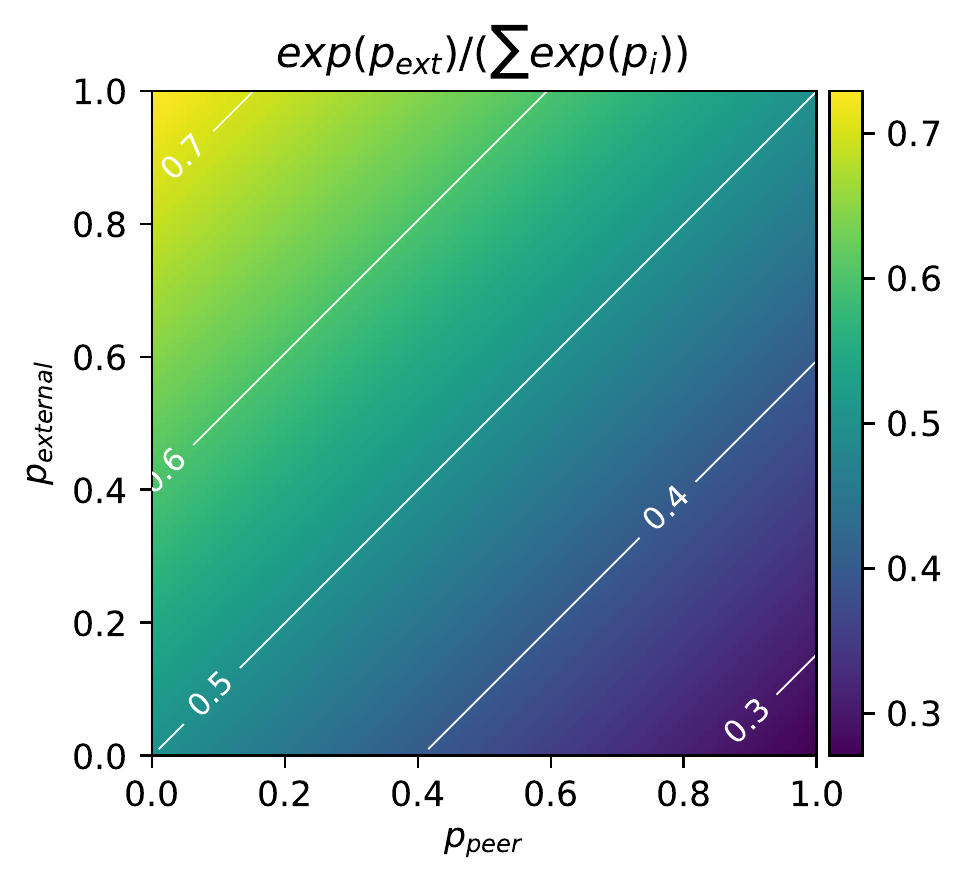}
    \caption{}
    \label{fig:responsibility_softmax}
  \end{subfigure}
  \begin{subfigure}[c]{0.33\textwidth}
    \includegraphics[width=1.0\linewidth]{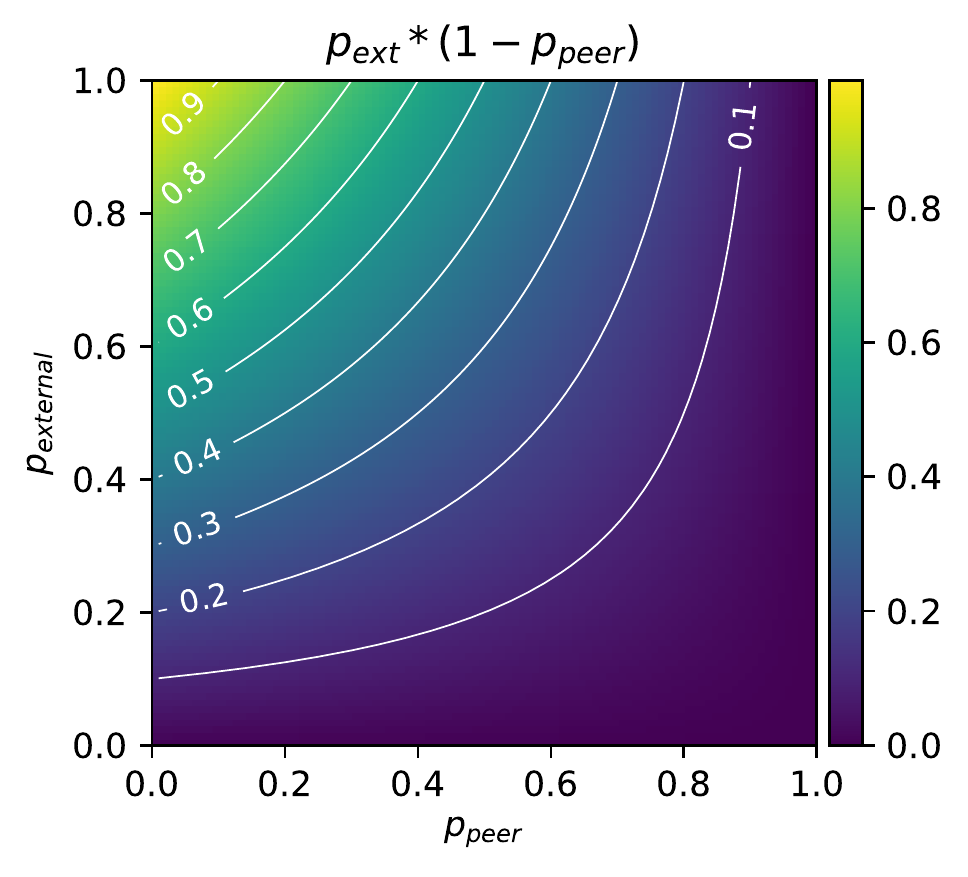}
    \caption{}
    \label{fig:responsibility_multiply}
  \end{subfigure}
  \caption{\textbf{Different definitions of exogenous responsibility.} Inference on simulated activation cascades under SI model with different definitions of exogenous responsibility. Qualitatively they all satisfy the same requirement - the larger the exogenous activation probability $p_{ext}$ the larger the exogenous responsibility. But only the definition in Figure~\ref{fig:responsibility} (Equation~\ref{eq:responsibility}) has a property that the high values of responsibility can be reached with relatively low values of $p_{ext}$. This makes sense in our case because we want our exogenous responsibility to reflect \emph{relative} difference between $p_{ext}$ and $p_{peer}$ rather than absolute difference. The AUC on simulated activation cascade is $0.92$ when using definition from Equation~\ref{eq:responsibility} (Figure~\ref{fig:responsibility}), $0.88$ when using definition from Equation~\ref{eq:responsibility_softmax} (Figure~\ref{fig:responsibility_softmax}), and $0.89$ when using definition from Equation~\ref{eq:responsibility_multiply} (Figure~\ref{fig:responsibility_multiply}). This indicates that the definition in Equation~\ref{eq:responsibility} is indeed the most appropriate.}
  \label{fig:different_exogenous_responsibility}
\end{figure}

Figure \ref{fig:different_exogenous_responsibility} shows experiments with SI model of endogenous influence but with three different definitions of external responsibility $R^{(i)}(t)$ which summarizers information in probability of endogenous activation $p^{(i)}_{peer}(t)$ at time $t$ for each user $i$ and probability of exogenous activation $p_{ext}(t)$ at time $t$, which is equal for all users. First, the original formulation which we use in the main paper:

\begin{equation}
R^{(i)}(t) = \frac{p_{\text{ext}}(t)}{p_{\text{ext}}(t) + p^{(i)}_{\text{peer}}(t)}
\label{eq:responsibility}
\end{equation}

Second, the softmax version of the measure above: 

\begin{equation}
R^{(i)}_{\text{softmax}}(t) = \frac{\exp(p_{\text{ext}}(t))}{\exp(p_{\text{ext}}(t)) + \exp(p^{(i)}_{\text{peer}}(t))}
\label{eq:responsibility_softmax}
\end{equation}

And third, the probability that you activated due to exogenous influence but \emph{not} due to endogenous influence: 

\begin{equation}
R^{(i)}_{\text{multiply}}(t) = p_{\text{ext}}(t) (1-p^{(i)}_{\text{peer}}(t))
\label{eq:responsibility_multiply}
\end{equation}

Qualitatively, they all satisfy a same requirement - the larger the $p_{\text{ext}}$ is, the larger the exogenous responsibility. However, only the definition in Equation~\ref{eq:responsibility} allows for the small values of $p_{\text{ext}}$ to achieve high exogenous responsibility, which is what we need because it is the \emph{relative} difference between $p_{\text{ext}}$ and $p^{(i)}_{\text{peer}}(t)$ that should influence the responsibility and not their absolute values. Plots on the right side of Figure~\ref{fig:different_exogenous_responsibility} illustrate this property visually.

\section{Inference on empirical datasets (extended)}

\begin{figure}[ht]
  \centering
  \includegraphics[width=\textwidth]{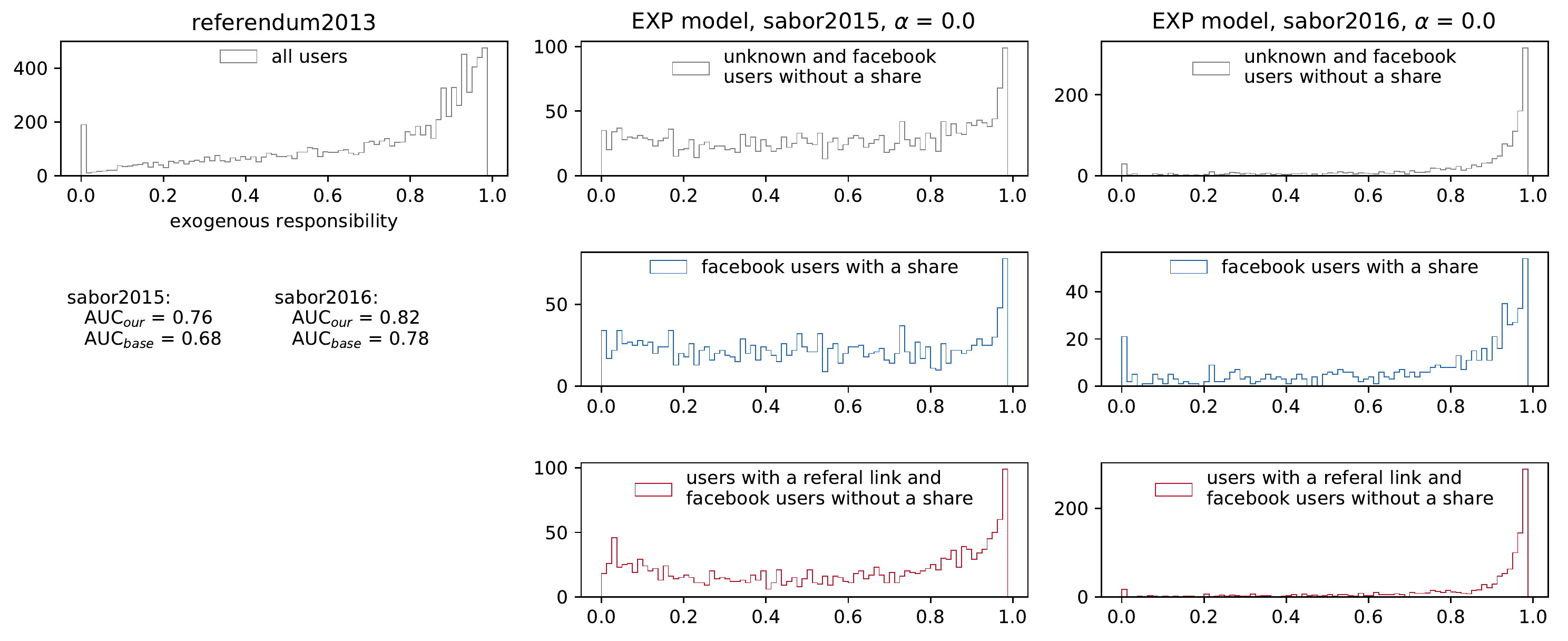}
  \caption{\textbf{Full distributions of exogenous responsibility in empirical datasets.} Histograms of exogenous responsibilities for all users which registered in one of three online survey applications. The three rows show all registered users (top), endogenously registered users (middle) and exogenously registered users (bottom). These measures are derived from the referral links from which users visited our application. For referendum2013 users we did not collect referral links so we cannot divide them into groups. We observe that exogenously registered users tend to have larger values of external responsibility in comparison to the endogenously registered users - their responsibility distributions are peaked more sharply near the high values. From these we can calculate the AUC scores to evaluate the performance of external responsibility as a classification measure. The achieved AUC scores ($AUC_{our}$) in this particular case when no correction for the observer bias was applied ($\alpha=0$) and where EXP is used as an endogenous influence model are $0.76$ and $0.82$ for the sabor2015 and sabor2016 datasets respectively. We also report AUC scores for the baseline ($AUC_{base}$) where we use the number of active peers as a measure of classifying users into endogenously and exogenously activated - $0.68$ and $0.78$ for the sabor2015 and sabor2016 datasets resectively, which is lower than we achieve with our method.}
  \label{fig:real_datasets_external_responsibility}
\end{figure}

Our methodology is able to characterize each activation on a scale from \emph{endogenous} to \emph{exogenous} depending on the relative magnitudes of endogenous and exogenous activation probabilities for each user. Figure \ref{fig:real_datasets_external_responsibility} shows histograms of exogenous responsibility for each user in each of the three datasets and the corresponding AUC scores which evaluate the quality of classification. As for sabor2015, and sabor2016, we actually have referral links from which users visited our online application we can use those as a gold standard for annotating the activations. Users that came to our application through referral link which corresponds to a share from Facebook could be safely labeled as endogenous activations. On the other hand, users who followed referral links which correspond to an external news site could be labeled as exogenous influence. We expect that users that activated due to exogenous influence have, on average, higher values of exogenous influence and so the weight of the distributions should be centered near the high values, while for endogenously activated users the distributions are more uniform in nature. This is a behavior that we observe in figure \ref{fig:real_datasets_external_responsibility}.

\begin{figure}[ht]
\centering
\includegraphics[width=0.8\linewidth]{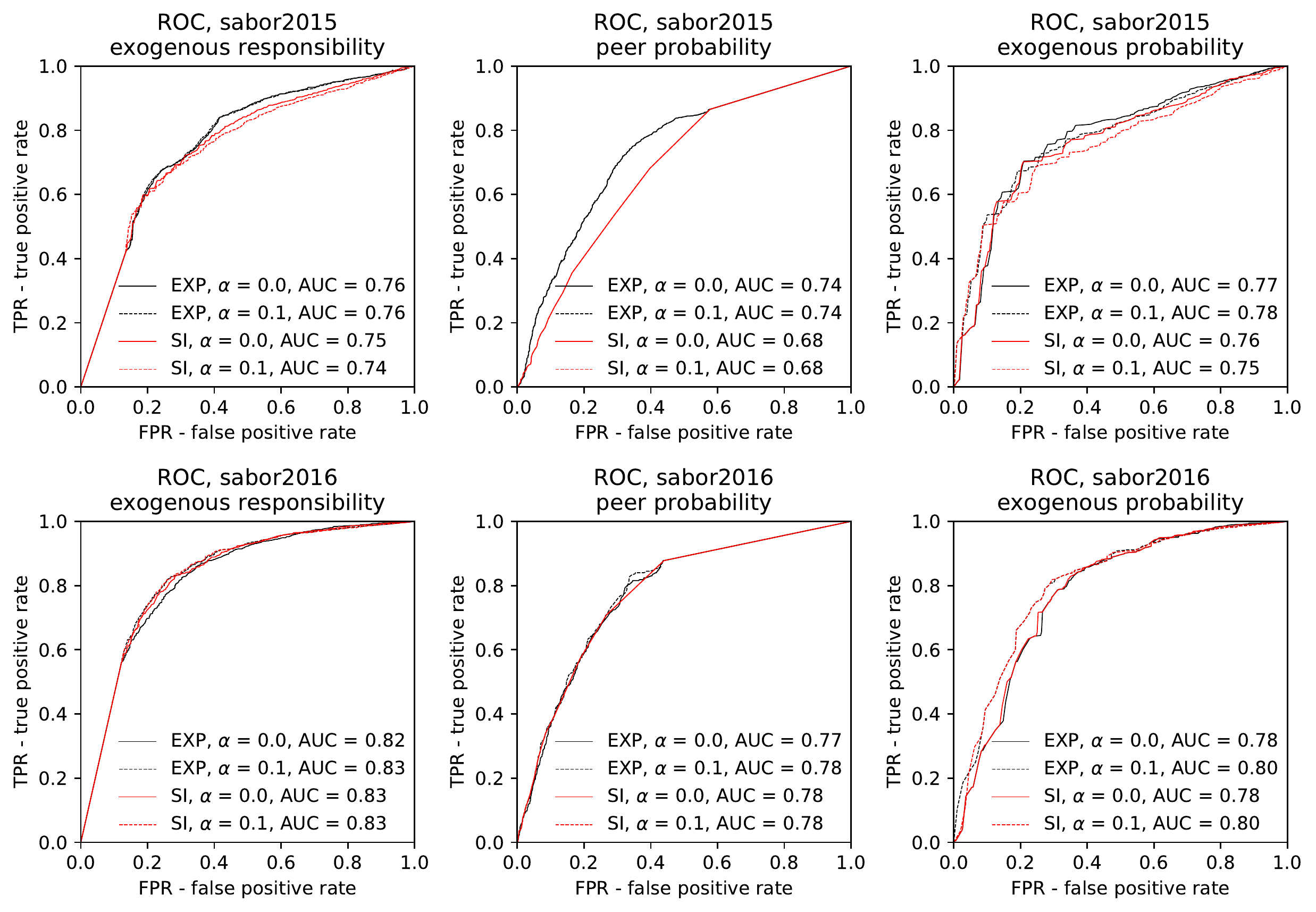}
\caption{\textbf{Different measures of exogenous influence.} Comparison of several possible measures of exogenous (external) influence: (i) endogenous (peer) activation probability (middle row), (ii) exogenous (external) activation probability (bottom row) (iii) external responsibility which is a combination of the two (top row). We performed experiments assuming both SI and EXP endogenous influence models, with and without correction for the observer bias.}
\label{fig:real_datasets_roc_curves_comparison}
\end{figure}

Next, we perform extensive experiments to test the utility of external responsibility as a measure of exogenous and endogenous influence. Looking at the Equation ~\ref{eq:responsibility} we see that we can equally likely use exogenous (external) activation probability $p^{(i)}_{\text{ext}}$ as well as endogenous (peer) activation probability $p^{(i)}_{\text{peer}}$. Figure~\ref{fig:real_datasets_roc_curves_comparison} shows the results of performing the inference on three datasets for which we have data on referral links - sabor2015 and sabor2016. We perform inference with two different endogenous influence models - SI and EXP model, and two different values of correction factor for observer bias (more details on the observer bias in section \ref{sec:observer_bias}): (i) without correction $\alpha=0$ and (ii) with correction $\alpha=0.1$. Although external responsibility does not achieve consistently best results on all datasets - exogenous (external) activation probability is slightly better on sabor2015 dataset, we believe it is still the most suitable as it achieves most consistent results on the simulated dataset (section~\ref{sec:simulated}, and Figure~\ref{fig:exogenous_activation_probability}).

\begin{figure}[ht]
  \centering
  \includegraphics[width=\textwidth]{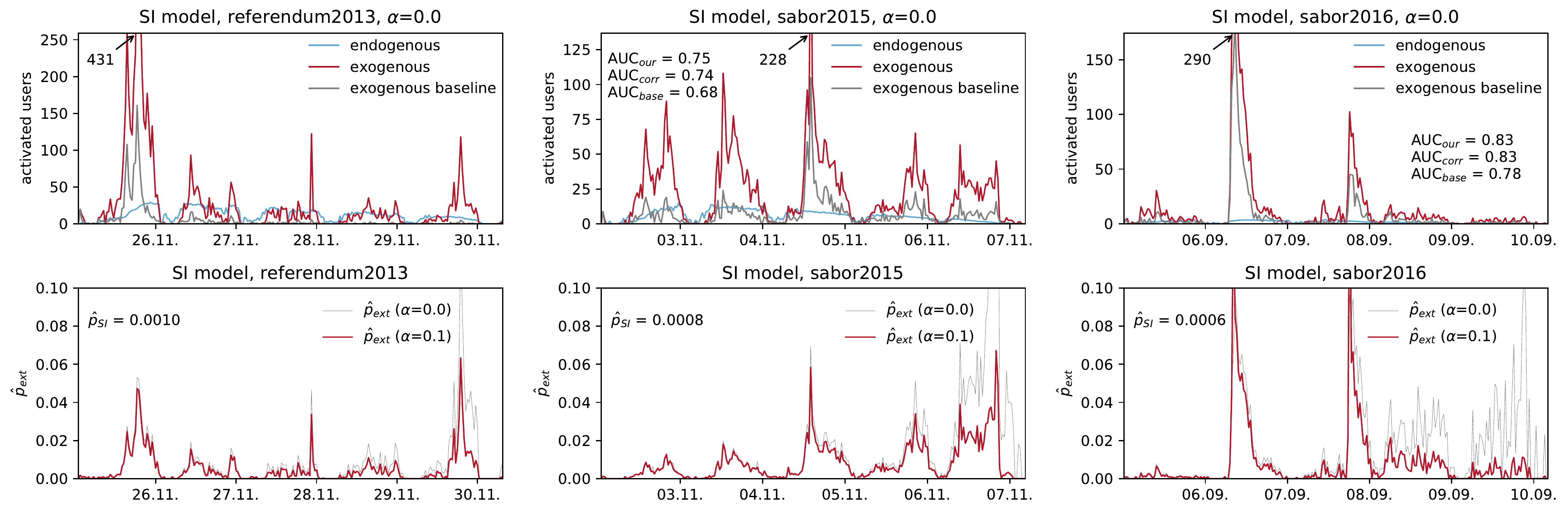}
\caption{\textbf{Inference on Facebook activation cascades with SI model.} Inference of endogenous and exogenous influence on activation cascades derived from referendum2013, sabor2015 and sabor2016 online survey applications, with SI as assumed endogenous influence model. Referrals originating from Facebook shares are diminishing in time, which closely resembles diminishing in endogenously activated users as estimated by our method. AUC scores for exogenous responsibility measure which classifies users into endogenous and exogenous activated show satisfying level of predictive performance. AUC score for sabor2015 and sabor2016 datasets for our method ($AUC_{our}$) are $0.75$ and $0.83$ respectively, which is higher as compared to baseline ($AUC_{base}$) which is $0.68$ and $0.78$ respectively. Facebook referrals alone are not discriminating enough as there are multiple possible ways by which Facebook users might reach our application, including visiting the web page of our application directly or through an advertisement, both which are more similar to exogenous rather than endogenous influence.}
\label{fig:real_datasets_results}
\end{figure}

\begin{figure}[ht]
  \centering
  \includegraphics[width=\textwidth]{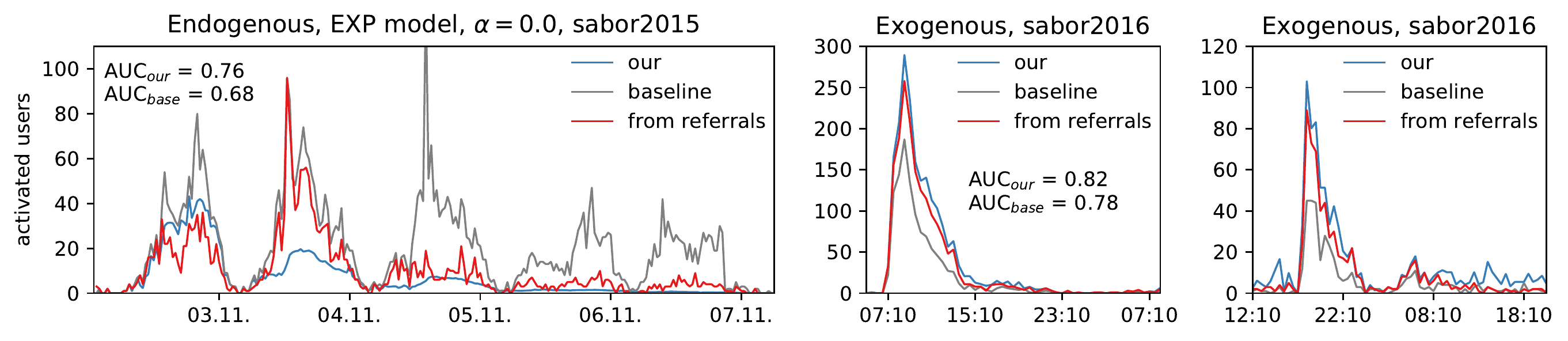}
  \includegraphics[width=\textwidth]{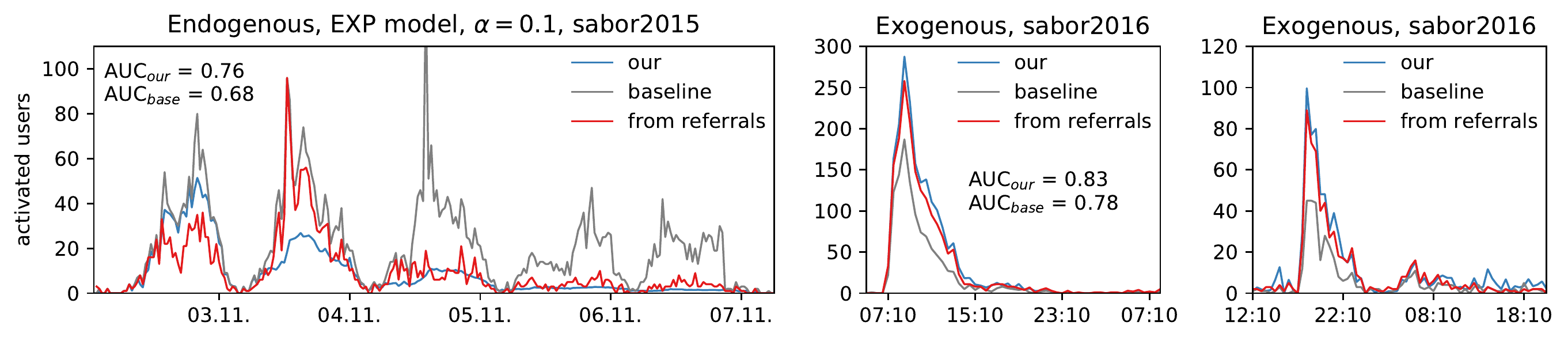}
\caption{\textbf{Inference on Facebook activation cascades with EXP model - comparison with baseline.} 
Comparison with the time series of activations obtained from user's referral links for sabor2015 and sabor2016 datasets shows that referrals originating from Facebook shares are diminishing in time (red line labeled with ``from referrals''), which closely resembles diminishing in endogenously activated users as estimated by our method (blue line labeled with ``our''). Achieved AUC scores in case of no correction for the observer bias ($\alpha=0$) with using exogenous responsibility as the criteria for classification of users into endogenously and exogenously activated are $0.76$ and $0.82$ for sabor2015 and sabor2016 datasets respectively, which is higher than the baseline where we use the number of active peers as the criteria instead, which achieves AUC scores of $0.68$ and $0.78$. Effect of applying correction for the observer bias ($\alpha=0.1$) is minor and raises AUC for sabor2016 dataset only from $0.82$ to $0.83$. This is probably due to the fact that the correction is strongest at the end of the observation period when there is less and less user activated due to the endogenous influence, which we approximate by observing which users followed referral links originating from a Facebook share. The effect of diminishing correction for the observer bias is also visible in Figure~\ref{fig:correction_observer_bias}.
}
\label{fig:real_datasets_baseline_comparison}
\end{figure}

\clearpage
\bibliography{bibliography-supporting}